\documentclass[twocolumn,floats,showpacs,superscriptaddress,pre]{revtex4}
\usepackage{amsfonts,amsmath,amssymb,hyperref}
\usepackage{graphicx}

\newcommand{\sample}{\hookleftarrow}
\newcommand{\h}{\mathbf{h}}
\newcommand{\R}{\mathbb{R}}
\newcommand{\G}{\mathcal{G}}

\newcommand{\arccosh}{\mathrm{arccosh}}

\begin{document}

\title{Duality between equilibrium and growing networks}

\author{Dmitri Krioukov}

\affiliation{Cooperative Association for Internet Data Analysis, University of California San Diego, CA, USA}

\author{Massimo Ostilli}

\affiliation{Cooperative Association for Internet Data Analysis, University of California San Diego, CA, USA}

\begin{abstract}
In statistical physics any given system can be either at an equilibrium or away from it. Networks are not an exception. Most network models can be classified as either equilibrium or growing. Here we show that under certain conditions there exists an equilibrium formulation for any growing network model, and vice versa. The equivalence between the equilibrium and nonequilibrium formulations is exact not only asymptotically, but even for any finite system size. The required conditions are satisfied in random geometric graphs in general and causal sets in particular, and to a large extent in some real networks.
\end{abstract}

\maketitle

\section{Introduction}

Statistical physics studies equilibrium and nonequilibrium systems using different theory and methods, as these systems are drastically different. Networks are not an exception. The vast realm of network models can be roughly divided into equilibrium and nonequilibrium domains~\cite{DorMen02a,Newman10-book}. In the former, one studies equilibrium ensembles of graphs of a fixed size. Classic examples are classical random graphs~\cite{SoRa51}, or soft configuration model and hidden variable models~\cite{BoPa03}. In nonequilibrium models, graphs grow, usually by adding nodes one at a time, introducing statistical dependencies. Preferential attachment~\cite{KraReLe00,DoMeSa00} is perhaps the best known example. These two approaches are clearly different~\cite{DorMen02a,BoPa03,BiBu03,BiBu05}. In the simplest example, i.e.\ classical random graphs ${\cal G}_{N,p}$, each pair of $N$ nodes are independently connected with the same probability $p$. The resulting degree distribution is Poissonian with mean $\bar{k}=pN$. In a growing version of this model, new nodes $N=1,2,3,\ldots$ are coming one at a time connecting to each existing node with probability $p=\bar{k}/N$. The degree distribution is exponential~\cite{DorMen02a}. The two ensembles are thus different since they generate graphs with different degree distributions.

In general, equilibrium systems tend to be more amendable for exact analytic treatment due to their simplicity. Networks are not an exception in this respect either, with examples including such powerful methods as exponential random graphs and other graph entropy tools~\cite{PaNe04,GaLo09,SqGa11,Bianconi08,BiPi09,AnBi09,AnBi11,ZhHa11,ZhKa11,WeBi12}. Exponential random graphs are particularly interesting from a theoretical perspective as they establish a precise connection between equilibrium ensembles of graphs, and canonical equilibrium ensembles in statistical mechanics. Yet the applicability of these equilibrium tools to real networks is questioned by the fact that real networks are not at an equilibrium; they are growing.

Here we show that if certain conditions are satisfied, then there exists a static (equilibrium) formulation ${\cal G}_S$ for any dynamic (growing) graph model ${\cal G}_D$, and vice versa. Specifically we prove that graph ensembles ${\cal G}_S$ and ${\cal G}_D$ are identical, and that this equivalence is exact not only asymptotically, but even for any finite graph size. That is, if ${\cal G}_S$ generates graph $G$ with probability $P(G)$, then so does ${\cal G}_D$.

We first discuss the required conditions in general (Section~\ref{sec:general}), explain why they are not satisfied in some popular network models (Section~\ref{sec:non-dual}), and provide examples of network models where these conditions are satisfied (Section~\ref{sec:dual}). These examples include random geometric graphs~\cite{Penrose03-book} in general and causal sets~\cite{BoLe87} in particular. The latter are random geometric graphs in Lorentzian spaces, and they were introduced as an approach to quantum gravity~\cite{BoLe87}. Causal sets in de Sitter spacetimes, such as the spacetime of our accelerating universe~\cite{Perlmutter98,Reiss98}, have been recently shown to model adequately some structural and dynamical properties of some real networks~\cite{KrKi12,PaBoKr11}, motivating our focus on random geometric graphs. In Section~\ref{sec:exact} we narrow down our consideration to those, and work out the details of equilibrium ${\cal G}_S$ and growing ${\cal G}_D$ models for random geometric graphs, proving the exact equivalence between these models, and confirming this equivalence in simulations.

\section{General duality conditions}\label{sec:general}

Random graphs with hidden variables~\cite{BoPa03} are a very general framework to which many popular network models belong as particular cases. The equilibrium ensemble of graphs ${\cal G}_S$ of size $N$ in this framework is defined in two steps:
\begin{enumerate}
\item for each node $t=1,2,\ldots,N$, sample its hidden variable $h_t$ from distribution $\rho_N(h)$, shorthand $h_t \sample \rho_N(h)$, and
\item connect each node pair $\{s,t\}$, $s<t\leq N$, with probability $p_N(h_s,h_t)$.
\end{enumerate}
That is, the ensemble of discrete states, i.e.\ graphs, is fully defined in terms of two continuous functions, the hidden variable probability density function (PDF) $\rho_N(h)$, and the connection probability $p_N(h,h')$.

The generalization to the growing case is straightforward. A sequence of growing graphs ${\cal G}_D$ of increasing size $t=1,2,\ldots$ is constructed by adding nodes numbered by $t$ one at a time, and:
\begin{enumerate}
\item for each new node $t=1,2,\ldots$, sample its hidden variable $h_t$ from distribution $\rho_t(h)$, and
\item connect new node $t$ to existing nodes $s$, $s<t$, with probability $p_t(h_s,h_t)$,
\end{enumerate}
where both hidden PDF and connection probability can in general depend on time or current graph size $t$.

The two ensembles ${\cal G}_S$ and ${\cal G}_D$ can be equivalent in the weak and strong senses. We say that they are equivalent in the {\em weak\/} sense if they generate graphs $G$ of size $N$ with the same probability $P(G)$. They are equivalent in the {\em strong\/} sense if this condition holds for graphs of {\em any\/} size. Denoting $G$'s adjacency matrix by $a_{st}$, $s,t=1,2,\ldots,N$, and by $p_{st}$ the probability of connection between nodes $s$ and $t$ in ${\cal G}_S$, $p_{st}=p_N(h_s,h_t)$, the probability that the equilibrium ${\cal G}_S$ construction generates graph $G$, given the hidden variable sampling $\{h_1,h_2,\ldots,h_N\}$, is
\begin{equation}\label{eq:P(G|h)}
P(G|h_1,h_2,\ldots,h_N) = \prod_{s<t}p_{st}^{a_{st}}\left(1-p_{st}\right)^{1-a_{st}}.
\end{equation}
Since all hidden variables $\{h_1,h_2,\ldots,h_N\}$ are independent, their joint PDF $\rho_N(h_1,h_2,\ldots,h_N)$ is the product of one-point PDFs,
\begin{equation}
\rho_N(h_1,h_2,\ldots,h_N) = \prod_{t=1}^N\rho_N(h_t),
\end{equation}
and the probability $P(G)$ of graph $G$ in the ${\cal G}_S$ ensemble is
\begin{equation}\label{eq:P(G)}
P(G) = \int P(G|\h)\rho_N(\h)\,d\h,
\end{equation}
where $\h=\{h_1,h_2,\ldots,h_N\}$. These equations imply that ${\cal G}_D$ is equivalent to ${\cal G}_S$ in the weak sense if the nonequilibrium ${\cal G}_D$ construction generates the same joint distribution of hidden variables $\rho_N(\h)$, and if the connection probabilities in both cases are also the same $p_N(h,h')$.

Indeed, if these two conditions hold, then the growing ${\cal G}_D$ definition for a graph of size $N$ is different from the equilibrium ${\cal G}_S$ definition only in the order in which node pairs $\{s,t\}$ are examined in the linking process, which is step~2 in both definitions. In the ${\cal G}_S$ case this order is manifestly random, induced by random labeling $t=1,2,\ldots,N$ of nodes, while in the ${\cal G}_D$ case this order appears to be preferred, induced by a preferred node labeling $t=1,2,\ldots,N$ reflecting node birth times. In reality however both orderings are random, and every node pair is examined once and connected or not connected with the same probabilities. Simply put, if the target graph size~$N$ is fixed in the growing case, then clearly one can set the connection probability in this case equal to the connection probability in the equilibrium case with the same~$N$, so that Eqs.~(\ref{eq:P(G|h)}-\ref{eq:P(G)}) apply to the growing ensemble ${\cal G}_D$ as well, although only in the weak sense, for the given~$N$. Since $P(G)$ is the same in the two ensembles, they are weakly equivalent. Since they are weakly equivalent, they generate {\em any\/} $N$-size graph $G$ with the same probability $P(G)$. Therefore the two sufficient conditions that $\rho_N(\h)$ and $p_N(h,h')$ are the same in both ensembles, are effectively the necessary conditions as well.

If the equilibrium and nonequilibrium ensembles are also strongly equivalent, then the connection probability cannot depend on time or graph size $N$, $p_N(h,h')=p(h,h')$. Indeed, suppose $h_1$ and $h_2$ are hidden variables of the first and second nodes in the growing ensemble ${\cal G}_D$. The probability $p_{12}$ of connection between the nodes can obviously depend only on $h_1$ and $h_2$ but not on $N$ simply by the definition of the strong equivalence implying that $p_{12}$ must be the same regardless of the target graph size $N$ until which the ${\cal G}_D$ growing process is let to run.

If the distribution of hidden variable does not depend on graph size either, $\rho_N(h)=\rho(h)$, then we have the simplest case of manifestly identical ${\cal G}_S$ and ${\cal G}_D$, but they generate dense graphs since the average degree in the ensembles is given by
\begin{equation}
\bar{k} = N\iint\rho(h)p(h,h')\rho(h')\,dh\,dh'.
\end{equation}
Almost all real networks are sparse. A graph ensemble is sparse if $\bar{k}=o(N)$ ($\bar{k}=O(1)$ or at most $\bar{k}=O(\log N)$ in most models of real networks). We thus see that for our graph ensembles ${\cal G}_S$ and ${\cal G}_D$ to be strongly equivalent and sparse, the distribution of hidden variables in them must depend on graph size. One possibility is $\rho_N(h)$ with a support that grows with $N$. Such $\rho_N(h)$ will have a normalization coefficient that decreases with $N$. We will see that this scenario is indeed enacted in random geometric graphs. But first we show why some popular network models do not satisfy the strong duality conditions considered above.

\section{Non-dual network models}\label{sec:non-dual}

In this section we consider three well-studied network models: classical random graphs, configuration model, and preferential attachment. The first two are equilibrium, and we show that there are no growing formulations that would be identical to these equilibrium ensembles in the strong sense. The last example is a growing network model, for which no strongly dual equilibrium formulation exists.

\subsection{Classical and regular random graphs}

Classical random graphs mentioned in the introduction provide perhaps the simplest example of strong non-duality of sparse graphs. This example belongs to the class of random graphs with hidden variables, except that there are no hidden variables, meaning that the hidden variable distributions (none) are the same in the equilibrium ${\cal G}_S$ and nonequilibrium ${\cal G}_D$ ensembles. However, the connection probability $p=\bar{k}/N$ manifestly depends on~$N$. Therefore for any given $N$, one has no problem defining ${\cal G}_D$ {\em weakly\/} equivalent to ${\cal G}_S$. Indeed, if one wishes to generate graph $G$ of size $N$ in the growing procedure with the same probability $P(G)$ as in the equilibrium procedure, one simply adds nodes $t=1,2,\ldots,N$ one at a time and connects them to existing nodes $s$, $s<t$, with constant probability $p=\bar{k}/N$. This way all node pairs are connected with the same probability as in the equilibrium procedure, except that node pairs are examined for linking not in a completely random but in some specific order.
We emphasize the  difference between this growing construction and the growing construction in the introduction where $N$ was not a constant target graph size, but the current growing graph size, ensuring that the average node degree in graphs of differen sizes in the ensemble was constant, i.e.\ did not depend on the graph size.

The only way to make the two ensembles {\em strongly\/} equivalent is to forget about average degree, and to connect each pair of nodes with the same constant probability $p$, making the two ensembles manifestly identical since both hidden variable distribution and connection probability are now the same and do not depend on $N$. In this case however, the resulting graphs are dense, and their average degree is $\bar{k}=Np$. The considerations above imply that there is no growing formulation that would be identical to the equilibrium ensemble for any graph size $N$ and fixed~$\bar{k}$.

The distribution of individual node degrees in classical random graphs is the Poisson distribution with mean $\bar{k}$, which is the maximum-entropy distribution with a given mean~\cite{Harremoes2001}. In that sense classical random graphs are maximum-entropy random graphs with a fixed expected node degree $\bar{k}$. But the node degrees can be fixed not only on average (colloquially, in the ``canonical ensemble'' or ``soft constraint'' sense), but also exactly to some integer $k$ (``microcanonical ensemble'' or ``hard constraints''). In the latter case we have $k$-regular random graphs---random graphs with all nodes having the same degree~$k$. The growing version of $k$-regular random graphs is manifestly impossible, since it is impossible to connect a $k$-degree node to a $k$-regular graph of size $N$, and obtain a $k$-regular graph of size $N+1$, because the $k$ existing nodes to which the new node connects will increase their degrees to $k+1$.

\subsection{Soft and hard configuration models}

The soft (canonical) and hard (microcanonical) configuration models are random graphs with a given sequence of expected degrees or exact degrees, respectively. To construct such graphs one usually first samples exact or expected degrees $k_t$, $t=1,2,\ldots,N$, from some target degree distribution~$\rho(k)$. In the soft case this distribution can be continuous, but in the hard case the sampled degrees must be integers. Upon such sampling, in the soft case, one connects all node pairs $\{s,t\}$ with probability $p_{st}=k_sk_t/(\bar{k}N)$---the soft model is thus a model of random graphs with hidden variables, where hidden variables $h_t$ are expected degrees $k_t$. In the hard case, one attaches $k_t$ edge stubs to nodes $t$, and then connects or matches random pairs of edge stubs attached to different nodes to form edges. We must emphasize here that these  constructions as described are not exactly correct. In the soft case, for example, the correct connection probability is not exactly $p_{st}=k_sk_t/(\bar{k}N)$~\cite{ChLu02} but $p_{st}=1/[1+\bar{k}N/(k_sk_t)]$~\cite{SqGa11,CoBo12}, and there are many details one has to worry about in the hard case~\cite{CoMa09,AnCo09,DeKi10}, but these details are not important for us here.

The important point is that in both soft and hard cases the probability of connections between nodes of (expected) degrees $k_s$ and $k_t$ scales as $p_{st} \sim k_sk_t/N$, i.e.\ it explicitly depends on $N$. Therefore considerations quite similar to those in the classical random graph case apply here as well, and no strongly dual growing formulations exist for these equilibrium ensembles either.

\subsection{Preferential attachment}

Preferential attachment is not a network model with hidden variables since the probability of connections depends not on any hidden variables but on observable degrees of nodes in a growing graph. However, considerations similar to previous examples apply here as well: since node degrees grow with the graph size, the connection probability depends on it as in those examples, violating the necessary condition for strong duality. As a consequence there can exist no strongly dual equilibrium formulation of the preferential attachment model.

A weakly dual formulation does exist as a mapping of preferential attachment to a hidden variable model in which the hidden variables are node birth times, and the connection probability is their function~\cite{BoPa03}. This model is not exactly identical to preferential attachment since the relation between the degrees of nodes and their birth times is not deterministic---the time-dependent distribution of degrees of nodes born at a given time is nontrivial and can be approximated by a Gamma distribution only in the thermodynamic limit~\cite{DorMen02a}. To the best of our knowledge there exists no weakly dual model that would be exactly identical to preferential attachment for finite graph sizes.

As a note closely related to the last point, the difficulties one must be prepared to face and deal with in the nonequilibrium settings are well illustrated by the fact that even though with some effort one can derive the exact expression for the distribution of node degrees in preferential-attachment graphs~\cite{KraReLe00,DoMeSa00}, it is impossible to write down a simple closed-form expression for probability $P(G)$ to generate graph $G$ in preferential attachment, cf.\ simple Eqs.~(\ref{eq:P(G|h)}-\ref{eq:P(G)}) in the equilibrium case.

\section{Dual network models}\label{sec:dual}

Random geometric graphs is a well-studied and perhaps the simplest network model for which strong duality holds. In this section we recall the definitions of random geometric graphs and causal sets, and provide a high-level explanation of why strong duality is possible in this case. The details of these dual definitions are worked out in the next section.

\subsection{Random geometric graphs}

Informally, random geometric graphs are discrete approximations of smooth geometries, with nodes representing ``atoms of space,'' and links representing some coarse information about proximity between these atoms. Formally, the equilibrium definition is: given a compact region in a geometric space, sprinkle a number of nodes uniformly at random over the region, and then connect each pair of nodes if the distance between them in the space is below a certain threshold. If the region grows somehow and so does the number of nodes, we have a growing network model.

For concreteness, we first consider the simplest case, the Euclidean disk of radius $R\gg1$. Its area or volume is $V=\pi R^2$. We want to sprinkle $N$ nodes into this disk. If we do so, the average node density in the disk will be $\delta=N/V$. If we sprinkle nodes uniformly, then any subarea of volume $dV$ will contain $dN=\delta\,dV$ nodes on average. To accomplish such uniform sprinkling in practice, one first selects a coordinate system---the polar coordinates $(r,\theta)$, $r\in[0,R]$, $\theta\in[0,2\pi]$, in the considered case, for example. In these coordinates, the metric, i.e.\ the square of the length of an infinitesimal line segment between points $(r,\theta)$ and $(r+dr,\theta+d\theta)$, is
\begin{equation}
ds^2=dr^2+r^2d\theta^2,
\end{equation}
while the volume form, i.e.\ the volume of an infinitesimal area between four points $(r,\theta)$, $(r+dr,\theta)$, $(r+dr,\theta+d\theta)$, and $(r,\theta+d\theta)$, is \begin{equation}\label{eq:dV-euc}
dV=r\,dr\,d\theta.
\end{equation}
Sprinkling nodes $t=1,2,\ldots,N$ onto the disk boils down to assigning to them their coordinates $(r_t,\theta_t)$. To ensure that sprinkling is uniform, $dN=\delta\,dV$, one has to respect the volume form~(\ref{eq:dV-euc}), which says that angular coordinates $\theta_t$ must be sampled from the uniform distribution on $[0,2\pi]$, $\theta_t\sample{\cal U}(0,2\pi)$, while the PDF $\rho(r)$ from which radial coordinates $r_t$ are sampled, $r_t\sample\rho(r)$, must be proportional to $r$, $\rho(r)=cr$, where $c=2/R^2$ is the normalization coefficient. Upon sampling the coordinates $(r_t,\theta_t)$ for all nodes $t=1,2,\ldots,N$ as described, one then connects all node pairs $\{s,t\}$ if the Euclidean distance $d_{st}$ between $s$ and $t$ is below a given threshold $d_c \ll R$. Each node $t$ thus connects to all other nodes that happen to lie within the disk of radius $d_c$ centered at $t$. Since the node density is uniform, the expected degree of all nodes, except those at $r_t\in[R-d_c,R]$, is the same and equal to $\bar{k}=\delta\pi d_c^2$. One can show that the distribution of node degrees is in fact Poisson with mean~$\bar{k}$.

This model is yet another example of equilibrium random graphs with hidden variables. Each node has two hidden variables, radial $r_t$ and angular $\theta_t$ coordinates distributed according to $\rho(r)$ and ${\cal U}(0,2\pi)$, and the connection probability is the Heaviside step function, $p_{st}=\Theta(d_c-d_{st})$. This connection probability does not depend on graph size, which is one necessary condition for strong duality. The other condition is that the distribution of hidden variables must be the same in the equilibrium and growing ensembles. The only way to satisfy this condition is to grow the size of a graph and the size of the disk that this graph occupies in a balanced way, such that at each step the joint distribution of node coordinates in the growing graph is the same as in the equilibrium case. Intuitively, it must be possible to satisfy this condition as well.

\begin{figure}
\includegraphics[width=3in]{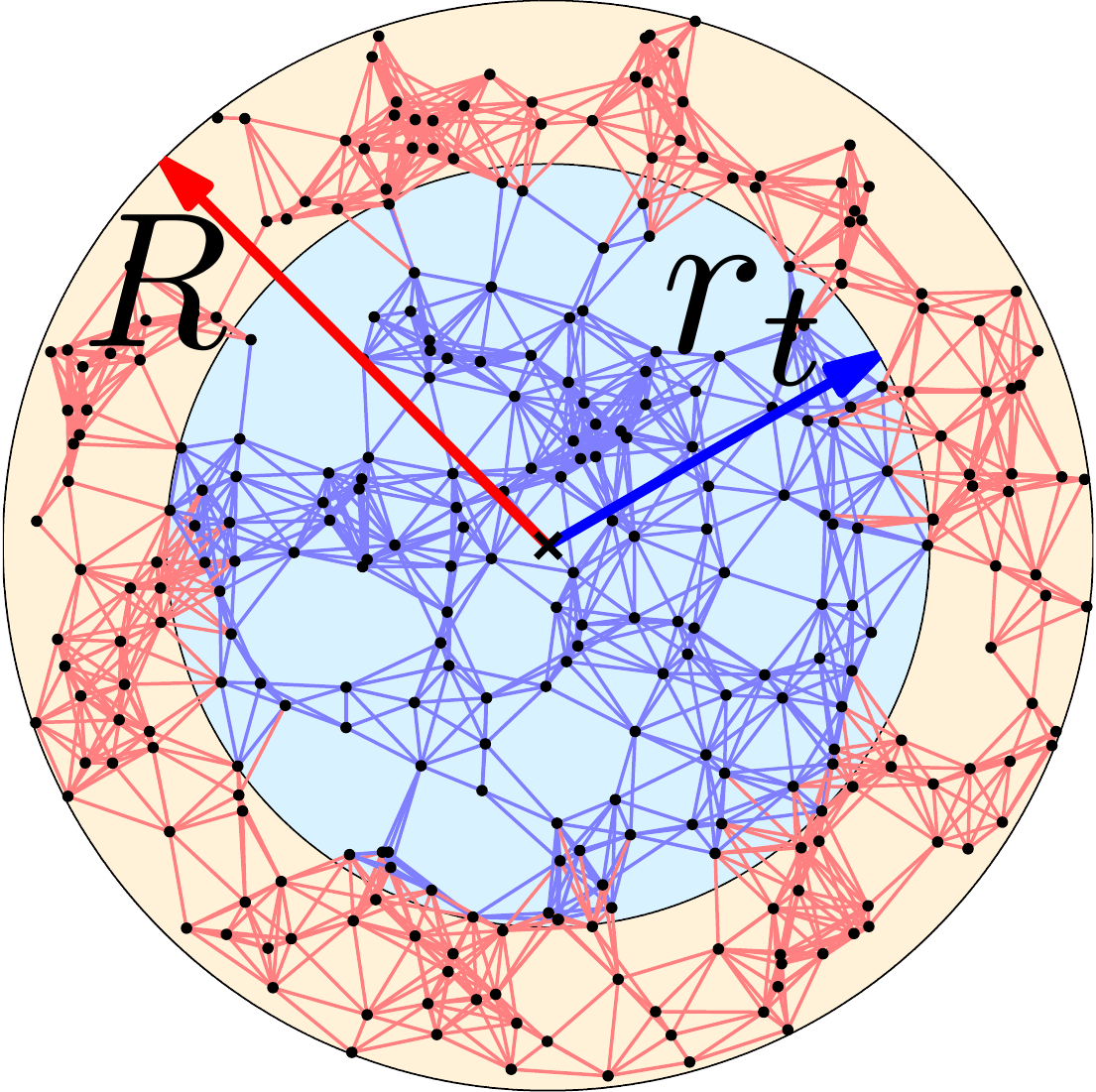}
\caption{(Color online) Random geometric graph $G$ on a Euclidean disk $D$ of radius $R$. Its subgraph $G_t$ is induced by all nodes lying in a smaller disk $D_t$ of radius $r_t<R$. ($N=300$, $R=1$, $r_t=0.7$, $d_c=0.2$.)
\label{fig:euc}}
\end{figure}

Indeed, consider a large equilibrium graph $G$ of size $N$ occupying disk $D$ of radius $R=\sqrt{N/(\delta\pi)}$, and consider the sequence of its subgraphs $G_t$, $t=1,2,\ldots,N$, induced by the $t$ nodes with the smallest radial coordinates $r_t$, $r_N=R$, see Fig.~\ref{fig:euc}. Intuitively, it is clear that each subgraph $G_t$ is also an equilibrium random geometric graph, except it is of smaller size $t\leq N$, occupying smaller disk $D_t$ of radius $r_t\leq R$. At the same time, we can consider sequence $G_t$, $t=1,2,\ldots,N$, as a growing graph occupying growing disk $D_t$. In fact, this growth process does not have to stop at $t=N$ and can continue forever. If new nodes $t$ connect to existing nodes with the same probability $p_{st}=\Theta(d_c-d_{st})$, then this growing ensemble is strongly equivalent to the equilibrium ensemble.

One has to implement this idea with care to make sure that at each step~$t$, the assignment of coordinates $(r_t,\theta_t)$ to new node~$t$ is such that the resulting joint distribution of the coordinates of all nodes $s=1,2,\ldots,t$ is exactly the same as in the equilibrium case, i.e.\ equal to the distribution of $t$ independent samples $\theta_s\sample{\cal U}(0,2\pi)$ and $r_s\sample 2r/r_t^2$. These implementation details making the statements above precise are in the next section.

Random geometric graphs can certainly be defined in spaces of any dimension and curvature, positive or negative. The main thing that changes is the metric. Changing metric changes the expression for distances between points, and the volume form. The former affects the distance calculations in the connection probability, while the latter defines the PDFs of node coordinates. The metric and volume form in the $d+1$-dimensional spaces, $d\geq1$, of curvature $+1$ (spherical space), $0$ (Euclidean space), and $-1$ (hyperbolic space), are, in spherical coordinates $(r,\theta_1,\ldots,\theta_d)$, $r\geq0$, $\theta_1,\ldots,\theta_{d-1}\in[0,\pi]$, $\theta_d\in[0,2\pi]$,
\begin{align}
ds^2&=dr^2+\sin^2r\,d\Omega_d^2,\quad &dV&=\sin^dr\,dr\,d\Phi_d,\label{eq:spherical_metric}\\
ds^2&=dr^2+r^2\,d\Omega_d^2,\quad &dV&=r^d\,dr\,d\Phi_d,\\
ds^2&=dr^2+\sinh^2r\,d\Omega_d^2,\quad &dV&=\sinh^dr\,dr\,d\Phi_d,\label{eq:hyperbolic_metric}
\end{align}
where
\begin{align}
d\Omega_d^2&=\sum_{i=1}^dd\theta_i^2\prod_{j=1}^{i-1}\sin^2\theta_j,
&d\Phi_d&=\prod_{i=1}^{d}\sin^{d-i}\theta_i\,d\theta_i
\end{align}
are the metric and volume form on the unit $d$-dimensional sphere. For example, if $d=2$, then $d\Omega_2^2=d\theta_1^2+\sin^2\theta_1\,d\theta_2^2$, which is the metric on the unit sphere, while its volume form is $d\Phi_2=\sin\theta_1\,d\theta_1\,d\theta_2$, so that the polar ($\theta_2$) coordinates of nodes must be sampled from ${\cal U}(0,2\pi)$, while their azimuthal ($\theta_1$) coordinates must be sampled from PDF $(1/2)\sin\theta$. If $d=1$ and the space is hyperbolic, then the PDF of radial coordinates $r$ on the disk of radius $R$ is $\sinh r/(\cosh R - 1)\approx\exp(r-R)$.

All the network models considered in Section~\ref{sec:non-dual} have asymptotically zero clustering. Contrary to those models, all random geometric graphs have finite clustering in the thermodynamic limit because triplets of nodes located close to each other in the space are all connected forming triangles. If the space is hyperbolic and if in addition the distance cutoff $d_c$ in the connection probability $p_{st}=\Theta(d_c-d_{st})$ is not constant but $d_c = R$, then the node degree distribution in resulting graphs is a power law with exponent $\gamma=3$~\cite{KrPa10}. However in this case there is no strong duality since the connection probability $p_{st}=\Theta(R-d_{st})$ depends on the graph size via disk radius~$R$.

\subsection{Causal sets}

Nothing stops us from extending the definition of random geometric graphs in Riemannian spaces considered above to pseudo-Riemannian spaces. The simplest and best-studied subcategory of pseudo-Riemannian spaces are Lorentzian spaces, and the simplest example of a Lorentzian space is the flat two-dimensional Minkowski spacetime. Its metric and volume form in coordinates $x,t\in\R$ are
\begin{equation}
ds^2=-dt^2+dx^2,\quad dV=dt\,dx.
\end{equation}
The problem that one immediately faces is that distances are no longer positive for distinct points, or zero if two points are the same. A pair of distinct points, called {\em events}, can now be separated by spacelike $\Delta s^2>0$, lightlike $\Delta s^2=0$, or timelike $\Delta s^2<0$ spacetime intervals $\Delta s$. Therefore if we want to define a random geometric graph in a Lorentzian space, then edges in this graph can no longer represent any meaningful proximity information---two faraway points in the Euclidean sense can be at zero distance in the Lorentzian sense. Consequently, no distance cutoff $d_c$ in the connection probability can make any sense any longer. Instead, the fundamental property of Lorentzian spaces is their causal structure. Two events are said to be causally related if they are timelike-separated, and the causal structure of a spacetime is nothing but the structure of these causal relations in it. The causal structure is a fundamental property of a Lorentzian spacetime because unless the spacetime is too pathological, its causal structure almost fully defines the spacetime itself~\cite{HaKi76,Malament77}.

\begin{figure}
\includegraphics[width=3in]{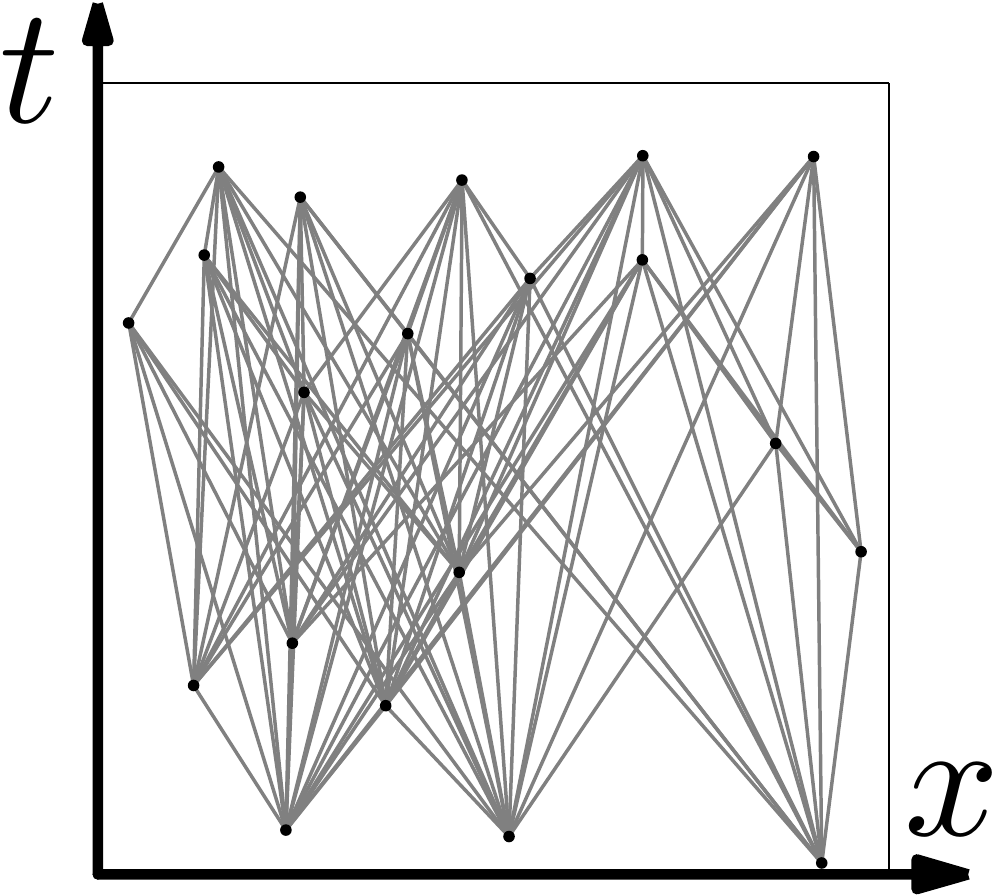}
\caption{Random geometric graph in box $x\in[0,R]$, $t\in[0,R]$ in two-dimensional Minkowski spacetime. ($N=20$, $R=1$.)
\label{fig:mink}}
\end{figure}

These considerations lead to the following equilibrium definition of random geometric graphs in Lorentzian spaces: given a compact region in a Lorentzian space, sprinkle a number of nodes uniformly at random over the region, and then connect each pair of nodes if they are timelike-separated. For example, given a box $x\in[0,R]$ and $t\in[0,R]$ in the two-dimensional Minkowski spacetime, we first sprinkle $N$ nodes $i=1,2,\ldots,N$ into the box by sampling
\begin{equation}
x_i,t_i\sample{\cal U}(0,R),
\end{equation}
and then connect each pair of nodes $\{i,j\}$, $i<j\leq N$, if
\begin{align}
\Delta x_{ij} &< \Delta t_{ij},\quad\text{where}\\
\Delta x_{ij} &= |x_i-x_j|,\\
\Delta t_{ij} &= |t_i-t_j|.
\end{align}
The sprinkling density is $\delta=N/R^2$. A sample graph is shown in Fig.~\ref{fig:mink}.

Random Lorentzian graphs defined above are called {\em causal sets\/} or {\em causets}, and they were proposed as an approach to quantum gravity~\cite{BoLe87}. Nodes in these causets are supposed to represent Planck-scale ``atoms'' of a quantum spacetime. One important task in the causet quantum gravity program is to derive some fundamental laws for causet growth dynamics that would grow causets similar to those obtained by equilibrium sprinkling onto the physical spacetime that we observe~\cite{RiSo99}. This difficult task is far from complete. If it succeeds one day, it might explain the emergence of the observable spacetime from some fundamental physical principles---a goal that other quantum gravity programs pursue as well. Although we show here that equilibrium causets can also be constructed in a growing fashion, to define this growth dynamics we must have a spacetime to start with.

Since the expansion of our universe is accelerating~\cite{Perlmutter98,Reiss98}, the spacetime that we live in is asymptotically de Sitter~\cite{HawkingEllis1975}. De Sitter spacetime has positive curvature, and it is the solution of Einstein's equation for an empty universe with positive cosmological constant~$\Lambda$, i.e.\ positive vacuum energy known as {\em dark energy}. In ``spherical'' coordinates $t\in\R$ (time) and $\theta_1,\ldots,\theta_{d-1}\in[0,\pi]$, $\theta_d\in[0,2\pi]$ (space), the metric and volume form in $d+1$-dimensional de Sitter spacetime of curvature $+1$ are
\begin{equation}\label{eq:de_sitter_metric}
ds^2=-dt^2+\cosh^2t\,d\Omega_d^2,\quad dV=\cosh^dt\,dt\,d\Phi_d.
\end{equation}
At each moment of time $t$, the spatial part of the spacetime (``current universe'') is thus a $d$-dimensional sphere of radius $\cosh t$.

If we consider only the half of de Sitter spacetime with $t\geq0$, and set $t \equiv r$, then the de Sitter equations~(\ref{eq:de_sitter_metric}) become evidently similar to the hyperbolic equations~(\ref{eq:hyperbolic_metric}). This similarity has been explored in~\cite{KrKi12}, where it was shown that de Sitter causets are asymptotically identical to the growing network model in~\cite{PaBoKr11}. This model is based on random geometric graphs growing in hyperbolic spaces, and it explains and accurately predicts not only many structural properties of some real networks, such as the Internet, social and biological networks, but also their growth dynamics. This asymptotic equivalence between the structure and dynamics of growing real networks, and the structure and dynamics of growing de Sitter causets, motivates our interest to the latter here.

\begin{figure}
\includegraphics[width=3in]{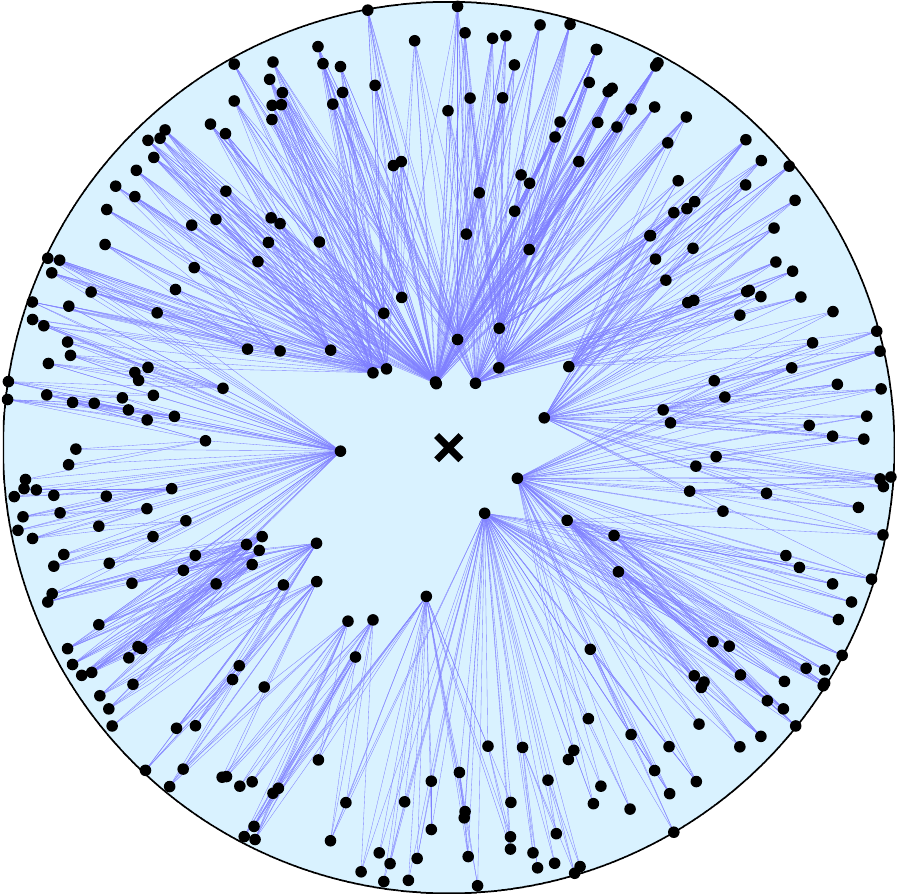}
\caption{(Color online) Random geometric graph in region $\eta\in[0,\eta_0]$ or $t\in[0,T]$, $\sec\eta_0=\cosh T$, in two-dimensional de Sitter spacetime. The angular coordinate of nodes are their spatial coordinates $\theta\in[0,2\pi]$, while their radial coordinates~$r$ are set equal to their temporal coordinates~$t$, $r=t=\arccosh\sec\eta$. ($N=300$, $T=5$.)
\label{fig:ds}}
\end{figure}

In practice it is often convenient to switch from the time coordinates $t\in\R$ to the conformal time coordinates $\eta\in(-\pi/2,\pi/2)$ defined by $\sec\eta=\cosh t$. In these conformal coordinates the metric and volume form become
\begin{equation}\label{eq:conformal_metric}
ds^2=\sec^2\eta\left(-d\eta^2+d\Omega_d^2\right),\quad dV=\sec^{d+1}\eta\,d\eta\,d\Phi_d,
\end{equation}
so that two events are timelike-separated if the conformal time difference $\Delta\eta$ between them exceeds their spatial distance $\Delta\Omega_d$. For example, if $d=1$, then
\begin{equation}
ds^2=\sec^2\eta\left(-d\eta^2+d\theta^2\right),\quad dV=\sec^{2}\eta\,d\eta\,d\theta.
\end{equation}
Therefore to build a causet of size~$N$ in a compact region of spacetime between, for instance, $\eta=0$ (``big bang'') and $\eta=\eta_0$ (``current time''), we first sample spatial $\theta_i$ and temporal $\eta_i$ coordinates for all nodes $i=1,2,\ldots,N$ from their PDFs defined by the volume form,
\begin{align}
\theta_i&\sample\rho(\theta)={\cal U}(0,2\pi),\\
\eta_i&\sample\rho(\eta)=\frac{\sec^2\eta}{\tan\eta_0},
\end{align}
and then connect all node pairs $\{i,j\}$, $i<j\leq N$, if
\begin{align}
\Delta\theta_{ij} &< \Delta\eta_{ij},\quad\text{where}\\
\Delta\theta_{ij}&=\pi-|\pi-|\theta_i-\theta_j||,\\
\Delta\eta_{ij}&=|\eta_i-\eta_j|.
\end{align}
The sprinkling density is $\delta=N/(2\pi\tan\eta_0)$. A sample graph is shown in Fig.~\ref{fig:ds}.

As proved in~\cite{KrKi12}, the graphs thus constructed have strongest possible clustering and power-law distribution of node degrees with exponent $\gamma=2$. This power law applies to nodes with degrees $k\gtrsim\delta$, i.e.\ to essentially all degrees in sufficiently large and sparse causets. Indeed, the causet size and average degree scale with the current time $T=\arccosh\sec\eta_0$ as $N\sim\delta e^T$ and $\bar{k}\sim\delta T$. Therefore, if $N$ is large and $\bar{k}$ is small, then $\delta$ is small, too. In denser causets with $\delta\gg1$, nodes of low degrees $k\lesssim\delta$ follow a power law with $\gamma=3/4$. Similar results apply to higher dimensions $d>1$ and plausibly to other spatial foliations because of the Lorentz invariance~\cite{KrKi12}. The higher the dimension, the weaker the clustering, but the distribution of node degrees $k\gtrsim\delta$ is still a power law with $\gamma=2$. The origin of this power law is a combination of two exponentials. The first exponential is the density of nodes~$\rho(t)$ as a function of their temporal coordinate~$t$, $\rho(t)\sim e^{dt}$. This exponential function is a direct consequence of the fact that the node density is uniform according to the volume form~(\ref{eq:de_sitter_metric}). The second exponential is the expected degree~$\bar{k}(t)$ of nodes born at time~$t$, $\bar{k}(t)\sim e^{-dt}$, proportional to the volume of their future light cones~\cite{KrKi12}. Combined, these two exponential scalings yield a power law with $\gamma=2$. As mentioned before, an analogous random geometric graph construction with the unform node density in hyperbolic spaces yields $\gamma=3$.

The important point is that the connection probability $p_{ij}=\Theta(\Delta\eta_{ij}-\Delta\theta_{ij})$ does not depend on the graph size or sprinkled area. Therefore, similar to Riemannian graphs, we expect strong duality to hold for these Lorentzian graphs as well. This is indeed the case as we show next.

\section{Exact duality}\label{sec:exact}

In this section we work out the definition details for the equilibrium $\G_S$ and growing $\G_D$ graph ensembles that are exactly identical to each other for any graph size. For concreteness and simplicity we limit our exposition to random geometric graphs in spaces with spherical symmetry considered in the previous section. In this case it is possible to derive these dual definitions from some basic facts about Poisson point processes (PPPs) on a positive real line.

\subsection{Reducing the problem to a PPP on $\R_+$}

Two graph ensembles with hidden variables are the same if the joint distributions of hidden variables in them are the same, and if their connection probabilities are also the same. In equilibrium and growing random geometric graphs, the connection probabilities are the same and do not depend on graph size, which is a necessary condition for strong duality. In random geometric graphs in spaces with spherical symmetry, the distribution of angular coordinates are also explicitly identical. In two dimensions, for example, the angular coordinates $\theta$ of all nodes are independent samples from the same distribution ${\cal U}(0,2\pi)$ in both equilibrium and growing formulations. Therefore the only remaining condition that must be satisfied for $\G_S$ and $\G_D$ to be strongly dual is that the joint distributions of radial coordinates in the Riemannian case or time coordinates in the Lorentzian case are also the same in $\G_S$ and $\G_D$.

The volume form in spaces with spherical symmetry can be written as
\begin{equation}
dV = f(x)\,dx\,d\Phi_d.
\end{equation}
In~(\ref{eq:spherical_metric}-\ref{eq:hyperbolic_metric}), for example, $x\equiv r$, and $f(x)=\sin^dx$, $f(x)=x^d$, and $f(x)=\sinh^dx$, respectively. In~(\ref{eq:de_sitter_metric}), $x \equiv t$, and $f(x)=\cosh^dx$, while in~(\ref{eq:conformal_metric}), $x=\eta$ and $f(x)=\sec^{d+1}x$. Therefore it is convenient to switch from the radial or temporal coordinates~$x$ to the volume coordinates~$v\in\R_+$ given by
\begin{align}
v &= \sigma_d\int_0^xf(x')\,dx',\quad\text{where}\label{eq:V(x)}\\
\sigma_d &= \int d\Phi_d = \frac{2\pi^\frac{d+1}{2}}{\Gamma\left(\frac{d+1}{2}\right)}
\end{align}
is the volume of the $d$-dimensional sphere. By the definition of these $v$-coordinates, if nodes are uniformly distributed in a compact region of space(time) with $x\in[0,X]$, then their $v$-coordinates are uniformly distributed on real interval $[0,V]$ in the usual sense, i.e.\ with the uniform PDF ${\cal U}(0,V)$, where $V=v(X)$.

The problem thus reduces to proper sampling of coordinates $v_i$ of nodes~$i$ on $\R_+$. Assuming we have a correct prescription for that, which is the subject of the rest of this section, to construct graph $G$ in $\G_S$ or $\G_D$, we then:
\begin{enumerate}
\item map $v_i$ to $x_i$ by inverting~(\ref{eq:V(x)}), which is similar to using the inverse CDF method to sample random variables $x_i$ from their PDF
\begin{equation}
x_i\sample\rho(x)=\frac{f(x)}{\int_0^Xf(x')\,dx'};
\end{equation}
\item sample the angular coordinates $\theta_1,\theta_2,\ldots,d\theta_d$ according to their PDFs defined by $d\Phi_d$; and
\item connect (new) node pairs with their connection probabilities.
\end{enumerate}
If the distribution of coordinates $v_i$ in $\G_S$ and $\G_D$ are the same, then the two ensembles are strongly dual.

We thus want to sample a (growing) number of points~$N$ on a (growing) real interval $[0,V]$ uniformly at random with density $\delta=N/V$. Uniform sampling implies that any subinterval $\Delta V$ contains $\Delta N=\delta\,\Delta V$ points on average. The simplest implementation of such uniform sampling is given by a Poisson point process (PPP) with rate~$\delta$. By definition, a PPP on a real line with rate~$\delta$ is a distribution of points on the line such that the number of points in any interval of length~$V$ is Poisson-distributed with mean~$\delta V$, and the numbers of points in disjoint intervals are independent random variables. In what follows we will rely on some basic textbook facts about PPPs~\cite{Snyder2011-book}.

\subsection{Equilibrium ensembles $\G_{S,V}$ and $\G_{S,N}$}

To properly define equilibrium ensembles for which dual growing ensembles would exist, we first observe that the prescription that we have followed so far, i.e.\ sampling exactly $N$ points from interval $I=[0,V]$ of a fixed length~$V$, cannot be exactly correct---$N$ and $V$ cannot be fixed simultaneously.

Indeed, let us fix $\delta=1$ and consider ensemble $\G_S$ on interval $[0,2]$ from which we sample exactly two points uniformly at random. In the growing ensemble $\G_D$ that we want to be identical to this $\G_S$, we first sample exactly one point from interval $[0,1]$ at the first graph-growing step, and then at the second step we sample exactly one point from interval $[1,2]$. Unfortunately the two ensembles cannot be identical because, for example, the number of points in interval $[0,1]$ in the described $\G_D$ is always~$1$, while the number of points in the same interval $[0,1]$ in $\G_S$ can be $0$ (both points happen to lie in $[1,2]$), $1$ (one point in $[0,1]$, the other in $[1,2]$), or $2$ (both points in $[0,1]$). The number of points in $[0,1]$ in $\G_S$ is equal to~$1$ only on average. Since the distributions of the number of points in an interval are not the same, then clearly the joint distribution of point locations cannot be the same either, and no duality is thus possible.

These observations demonstrate that sampling a fixed number of points from a fixed interval does not correctly implement our intention to ``sample points uniformly at random.'' The simplest correct implementation, a PPP with rate~$\delta$, can be formulated either for a fixed interval or for a fixed number of nodes, but not for both. That is, we cannot have just one equilibrium ensemble $\G_S$ of graphs of fixed size $N$ occupying a space(time) region of fixed volume~$V$. We necessarily have two different ensembles $\G_{S,V}$ and $\G_{S,N}$. In the former the space(time) volume is fixed to~$V$; in the latter the graph size is fixed to~$N$.

By the PPP definition, the correct procedure to sample coordinates $v_i$ of nodes in the fixed-volume $V$ ensemble $\G_{S,V}$ is:
\begin{enumerate}
\item sample graph size $N$ from the Poisson distribution
\begin{equation}\label{eq:poisson}
N\sample\mathbb{P}_{\delta V}(N) = e^{-\delta V}\frac{\left(\delta V\right)^N}{N!};
\end{equation}
\item sample $N$ random numbers $v_i$, $i=1,\ldots,N$, from the uniform distribution on $[0,V]$,
\begin{equation}
v_i\sample{\cal U}(0,V).
\end{equation}
\end{enumerate}
That is, $N$ cannot fixed. It must be a Poisson-distributed random variable. Its mean is
\begin{equation}
\bar{N}=\sum_{N=0}^\infty N\mathbb{P}_{\delta V}(N)=\delta V.
\end{equation}

The only complication with the fixed-graphsize $N$ ensemble $\G_{S,N}$ is that we have to know the distribution of the $v$-coordinate $V$ of the $N$'th point in a PPP. This distribution is also known as the distribution of waiting times $V$ for the $N$'th outcome in a Poisson process, and it is given by the Gamma distribution. Therefore the correct procedure to sample $v_i$, $i=1,2,\ldots,N$, in $\G_{S,N}$ is:
\begin{enumerate}
\item sample space(time) volume $V$ from the Gamma distribution dual to the Poisson distribution $\mathbb{P}_{\delta V}(N)$
\begin{equation}\label{eq:gamma}
V\sample\Gamma_{N,\delta}(V) = e^{-\delta V}\frac{\left(\delta V\right)^N}{N!}\frac{N}{V}
\end{equation}
\item set $v_N=V$;
\item for $i=1,2,\ldots,N-1$, sample
\begin{equation}
v_i\sample{\cal U}(0,V).
\end{equation}
\end{enumerate}
That is, $V$ cannot fixed. It must be a Gamma-distributed random variable. Its mean is
\begin{equation}
\bar{V}=\int_{V=0}^\infty V\Gamma_{N,\delta}(V)=\frac{N}{\delta}.
\end{equation}

Since the Gamma distribution $\Gamma_{N,\delta}(V)$ is the distribution of volumes $V$ occupied by $N$ points in a PPP with rate $\delta$, the $\G_{S,N}$ definition implements the same PPP as $\G_{S,V}$, except that not the volume $V$ but the number of points $N$ is now fixed.

\subsection{Growing ensembles $\G_{D,V}$ and $\G_{D,N}$}

Similar to the equilibrium case, in the growing case we can grow either volume $V$ or graph size $N$ by fixed increments, sampling the other variable from an appropriate distribution. We cannot increase both $N$ and $V$ by fixed amounts.

The simplest and most practically relevant case that we consider first is the growing ensemble $\G_{D,N}$ where at each time step $i=1,2,\ldots,N$ we add exactly one node $i$ as follows:
\begin{enumerate}
\item sample volume increment $V$ from the exponential distribution,
\begin{equation}
V\sample\Gamma_{1,\delta}(V)=\delta e^{-\delta V},
\end{equation}
\item assuming $v_0=0$, set the new node coordinate to
\begin{equation}
v_i = v_{i-1} + V.
\end{equation}
\end{enumerate}
This definition relies on the fact that the distribution of waiting times~$V$ for an outcome of a Poisson process with rate~$\delta$ is exponential, equal to the Gamma distribution with~$N=1$. The $v$-coordinate of new node $i$ is then equal to the $v$-coordinate of the previous node $i-1$, plus this waiting time~$V$. This construction implements the same PPP as in the previous subsection.

\begin{figure}
{\includegraphics[width=3.25in]{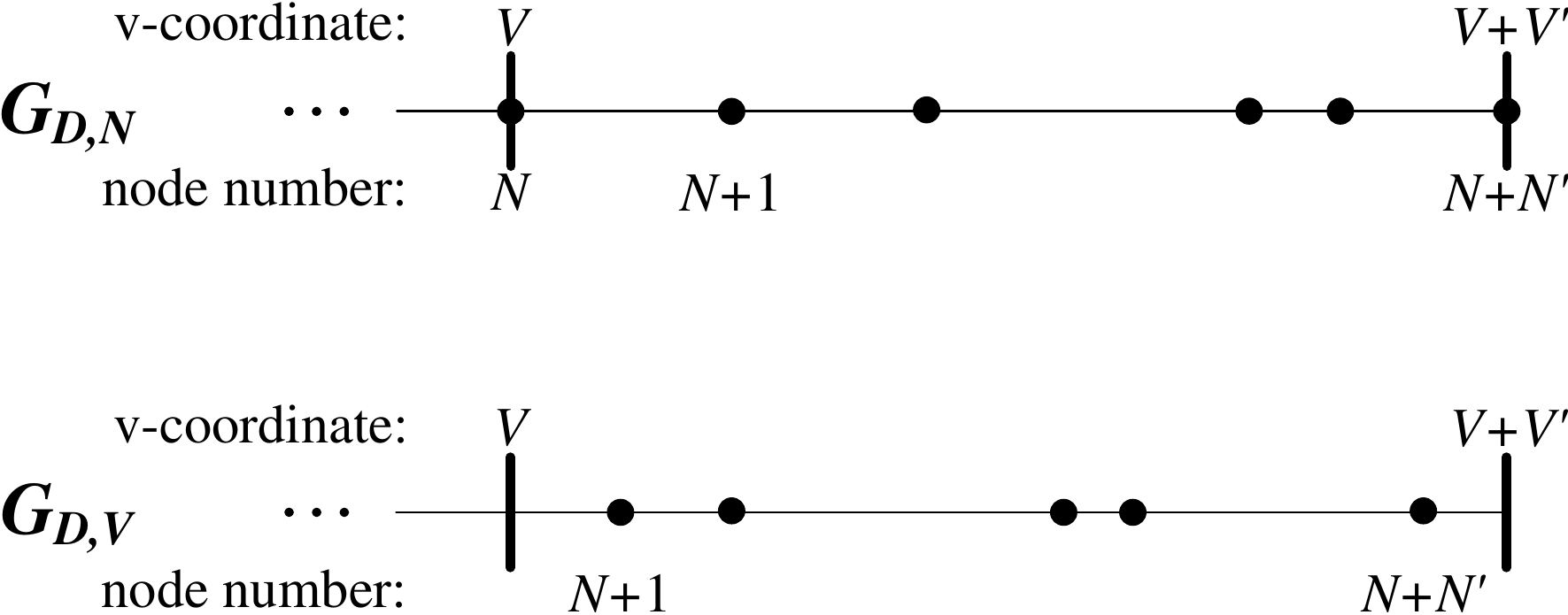}}
\caption{Illustration of the Poisson point process in $\G_{D,N}$ (top) and $\G_{D,V}$ (bottom).
\label{fig:0}
}
\end{figure}

We can also extend this $\G_{D,N}$ definition to a more general case where we add any fixed number of nodes at each graph-growing step. Denoting by $N$ the current graph size (before adding nodes), $V$ the last ($N$'th) node coordinate $v_N$, and $N'$ the number of new nodes to add, here is the general $\G_{D,N}$ definition, see Fig.~\ref{fig:0}(top):
\begin{enumerate}
\item sample volume increment $V'$ from the Gamma distribution,
\begin{equation}
V'\sample\Gamma_{N',\delta}(V'),
\end{equation}
\item set the coordinate of the new last ($N+N'$'th) node to $V+V'$,
\begin{equation}
v_{N+N'} = V+V',
\end{equation}
\item sample coordinates for the rest of new nodes $i=N+1,N+2,\ldots,N+N'-1$ from the uniform distribution on $[V,V+V']$,
\begin{equation}
v_i\sample{\cal U}(V,V+V').
\end{equation}
\end{enumerate}

The other option is $\G_{D,V}$ where we increase the volume by a fixed amount. This amount can be $1/\delta$ at each step, so that the {\em expected\/} number of nodes we add each time is~$1$. In general this volume increment can be any positive number. Denoting by $V$ and $N$ the current volume and graph size (before adding new nodes), and by $V'$ the volume increment, the graph-growing step in $\G_{D,V}$ is defined as follows, see Fig.~\ref{fig:0}(bottom):
\begin{enumerate}
\item sample the number of new nodes $N'$ from the Poisson distribution dual to $\Gamma_{N',\delta}(V')$,
\begin{equation}
N'\sample\mathbb{P}_{\delta V'}(N'),
\end{equation}
\item sample coordinates of new nodes $i=N+1,N+2,\ldots,N+N'$ from the uniform distribution on $[V,V+V']$,
\begin{equation}
v_i\sample{\cal U}(V,V+V').
\end{equation}
\end{enumerate}
The number of new nodes $N'$ can be zero with the larger probability, the smaller $V'$.

\subsection{Relationship between the four ensembles}

It is evident that the definitions of the graph-growing steps in $\G_{D,V}$ and $\G_{D,N}$ are identical to the definitions of equilibrium ensembles $\G_{S,V}$ and $\G_{S,N}$ applied to different intervals ($[V,V+V']$ in $\G_{D,V}$ instead of $[0,V]$ in $\G_{S,V}$) or node sets ($N+1,\ldots,N+N'$ in $\G_{D,N}$ instead of $1,\ldots,N$ in $\G_{S,N}$). All the four ensembles are thus nothing but four different implementations of the same PPP with rate~$\delta$ on $\R_+$.

In fixed-volume ensembles $\G_{S,V}$ and $\G_{D,V}$, we take a snapshot of this PPP on interval $[0,V]$. In fixed-graphsize ensembles $\G_{S,N}$ and $\G_{D,N}$, we take different a snapshot of the same PPP on node set $1,\ldots,N$. Therefore we have two pairs of identical ensembles, $\G_{S,V}=\G_{D,V}$ and $\G_{S,N}=\G_{D,N}$. Not only the ensembles in each pair are dual in the sense of exact equivalence, but the pairs themselves are also dual in the sense that they provide two dual fixed-$V$.vs.$N$ views on the same PPP. In the thermodynamic limit $N,V\to\infty$ we have the full view of the same PPP on the whole $\R_+$. The two pairs of ensembles converge to each other, i.e.\ all the four ensembles become identical.

In other words, we have the following diagram
\begin{equation*}
\begin{array}{ccc}
\G_{S,V}&=&\G_{D,V}\\
\wr&&\wr\\
\G_{S,N}&=&\G_{D,N}
\end{array}
\end{equation*}
where the equal sign `$=$' means the exact equivalence for any system size, while symbol `$\sim$' stands for the asymptotic equivalence and for the $V$.vs.$N$ PPP duality at finite sizes. The proofs of these statements follow directly from the fact that all these ensembles implement the same PPP. For completeness, we provide these proofs here.

\subsection{Duality proofs}

\subsubsection{Fixed-volume ensembles}

To show that the joint distributions of node coordinates in $\G_{S,V}$ and $\G_{D,V}$ are the same, it suffices to show this for the first two graph-growing steps in $\G_{D,V}$. Let $V,V'$ and $N,N'$ be the fixed volumes and Poisson-sampled numbers of nodes at the first and second steps,
\begin{align}
N&\sample\mathbb{P}_{\delta V}(N),\\
N'&\sample\mathbb{P}_{\delta V'}(N').
\end{align}
Sampling $N$ real numbers ($v$-coordinates) uniformly from $[0,V]$, and then sampling $N'$ numbers uniformly from $[V,V+V']$ is clearly identical to sampling $N+N'$ numbers uniformly from $[0,V+V']$, and the joint distributions of samples in both cases are the same.

In the $\G_{S,V}$ corresponding to the described $\G_{D,V}$, $N''$ numbers are sampled uniformly from the same interval $[0,V+V']$, where
\begin{align}
N''&\sample\mathbb{P}_{\delta(V+V')}(N'').
\end{align}
The joint distributions of $v$-coordinates are thus exactly the same in $\G_{S,V}$ and $\G_{D,V}$ if the probability that $N''=X$ is equal to the probability that $N+N'=X$, which is indeed the case because a sum of Poisson-distributed random variables is Poisson-distributed with the mean equal to the sum of means.

\subsubsection{Fixed-graphsize ensembles}

Similarly, to show that the joint distributions of node coordinates in $\G_{S,N}$ and $\G_{D,N}$ are the same, it suffices to show this for the first two nodes $i=1,2$ in $\G_{D,N}$. The generalization to the first two steps with arbitrary numbers of nodes is straightforward. We work with joint distributions $\rho(v_1,v_2)$ whose arguments are ordered $v_1\leq v_2$; if $v_1>v_2$, then $\rho(v_1,v_2)=0$. The unordered joint distribution is then $\rho^*(v_1,v_2)=[\rho(v_1,v_2)+\rho(v_2,v_1)]/2$, or in general $\rho^*(v_1,v_2,\ldots,v_N)=(1/N!)\sum_\pi\rho(v_{\pi(1)},v_{\pi(2)},\ldots,v_{\pi(N)})$, where the summation is over the $N!$ permutations $\pi$ of indices $1,2,\ldots,N$.

Let $V,V'$ be the first and second volume increments in $\G_{D,N}$ with adding one node at each step,
\begin{align}
V&\sample\Gamma_{1,\delta}(V)=\delta e^{-\delta V},\\
V'&\sample\Gamma_{1,\delta}(V')=\delta e^{-\delta V'}.
\end{align}
These distributions define the distribution of the first node coordinate $v_1$ and the conditional distribution of the second node coordinate $v_2$:
\begin{align}
\rho_D(v_1)&=\delta e^{-\delta v_1},\\
\rho_D(v_2|v_1)&=\delta\Theta(v_2-v_1)e^{-\delta(v_2-v_1)},
\end{align}
where we have inserted the Heaviside step function $\Theta(v_2-v_1)$ to emphasize that $v_1 \leq v_2$. The joint distribution is then
\begin{equation}
\rho_D(v_1,v_2)=\rho_D(v_2|v_1)\rho_D(v_1)=\delta^2\Theta(v_2-v_1)e^{-\delta v_2}.
\end{equation}

In the corresponding $\G_{S,N}$, the total volume $V''$ is sampled from $\Gamma_{2,\delta}(V'')$,
\begin{align}
V''&\sample\Gamma_{2,\delta}(V'')=\delta^2 V''\,e^{-\delta V''}.
\end{align}
The first node coordinate is sampled uniformly from $[0,V'']$, while the second node coordinate is equal to $V''$. In other words, the distribution of the second node coordinate $v_2$ and the conditional distribution of the first node coordinate $v_1$ are
\begin{align}
\rho_S(v_2)&=\delta^2 v_2e^{-\delta v_2},\\
\rho_S(v_1|v_2)&=\frac{\Theta(v_2-v_1)}{v_2}.
\end{align}
Completing the proof, the joint distribution is then
\begin{align}
\rho_S(v_1,v_2)&=\rho_S(v_1|v_2)\rho_S(v_2)=\delta^2\Theta(v_2-v_1)e^{-\delta v_2}\nonumber\\
&=\rho_D(v_1,v_2).
\end{align}

As a consequence, the marginal distributions for the first and second coordinates are identical in $\G_{S,N}$ and $\G_{D,N}$:
\begin{align}
\rho_S(v_1)&=\int_{v_2=0}^\infty\rho_S(v_1,v_2)\,dv_2=\Gamma_{1,\delta}(v_1)=\rho_D(v_1),\\
\rho_D(v_2)&=\int_{v_1=0}^\infty\rho_D(v_1,v_2)\,dv_1=\Gamma_{2,\delta}(v_2)=\rho_S(v_2).
\end{align}
The last equation is a particular instance of the general fact that the sum of Gamma-distributed random variables is Gamma-distributed with the mean equal to the sum of means.

\subsection{Importance of the joint distribution}

One may question our focus on the joint distributions above: do they really matter, and would not it be sufficient to show that the one-point coordinate densities in dual ensembles coincide? The equality between one-point distributions is a necessary but definitely not sufficient condition because one-point distributions are marginals of joint distributions, and two different joint distributions can have the same marginal. If the joint coordinate distributions are different, then even if their marginals are the same, some graph properties can be different, even in the thermodynamic limit. One example of such properties is the maximum degree. As one can see in Fig.~\ref{fig:ds}, the expected degree of a node is a decreasing function of its temporal coordinate in de Sitter causets. Therefore the distribution of the maximum degree depends on the distribution of the smallest time coordinate. The latter distribution is given be the order statistics~\cite{OrderStatistics03}, but the order statistics is fully determined only by the joint distribution, and not by its marginals. In general, two different joint distributions with the same marginals have different order statistics, and consequently different distributions of smallest or largest samples.

In that regard, it is instructive to consider the following fixed-$N$ ensemble $\G_{W,N}$, where `W' stands for ``wrong'': for all nodes $i=1,\ldots,N$, sample their coordinates
\begin{equation}
v_i\sample\Gamma_{i,\delta}(v_i).
\end{equation}
The erroneous intuition might be that since the $v$-coordinate of the $i$'th node in a PPP is distributed according to $\Gamma_{i,\delta}(v)$, then we can safely sample directly from this distribution, and one can check that if we do so, then the single-point PDF will be indeed equal to the one in $\G_{S,N}=\G_{D,N}$, derived below. However, this process is no longer the same PPP, and its joint PDF is different from the one in $\G_{S,N}=\G_{D,N}$, because the ordering of coordinates can now be violated. Indeed, $v_i$ can with certain probability be larger than $v_j$ for any $i<j$, while by definition, $\Gamma_{i,\delta}(v)$ is the distribution of the $i$'th largest coordinate in a PPP. As a result $\G_{W,N}$ is not identical to $\G_{S,N}=\G_{D,N}$ even in the thermodynamic limit, which one can verify in simulations by checking the maximum and average degree statistics, for instance.

\subsection{Simulations}

To confirm in simulations that we have constructed two pairs of identical graph ensembles $\G_{S,V}=\G_{D,V}$ and $\G_{S,N}=\G_{D,N}$, it would suffice to show that the joint distribution of node coordinates in the equivalent pairs are the same. However, since visualizing joint distributions of a large number of variables is impractical, we limit ourselves to showing their one-point PDFs. For concreteness, we do so for causal sets in $1+1$-dimensional de Sitter spacetime.

\begin{figure}
{\includegraphics[width=3.25in]{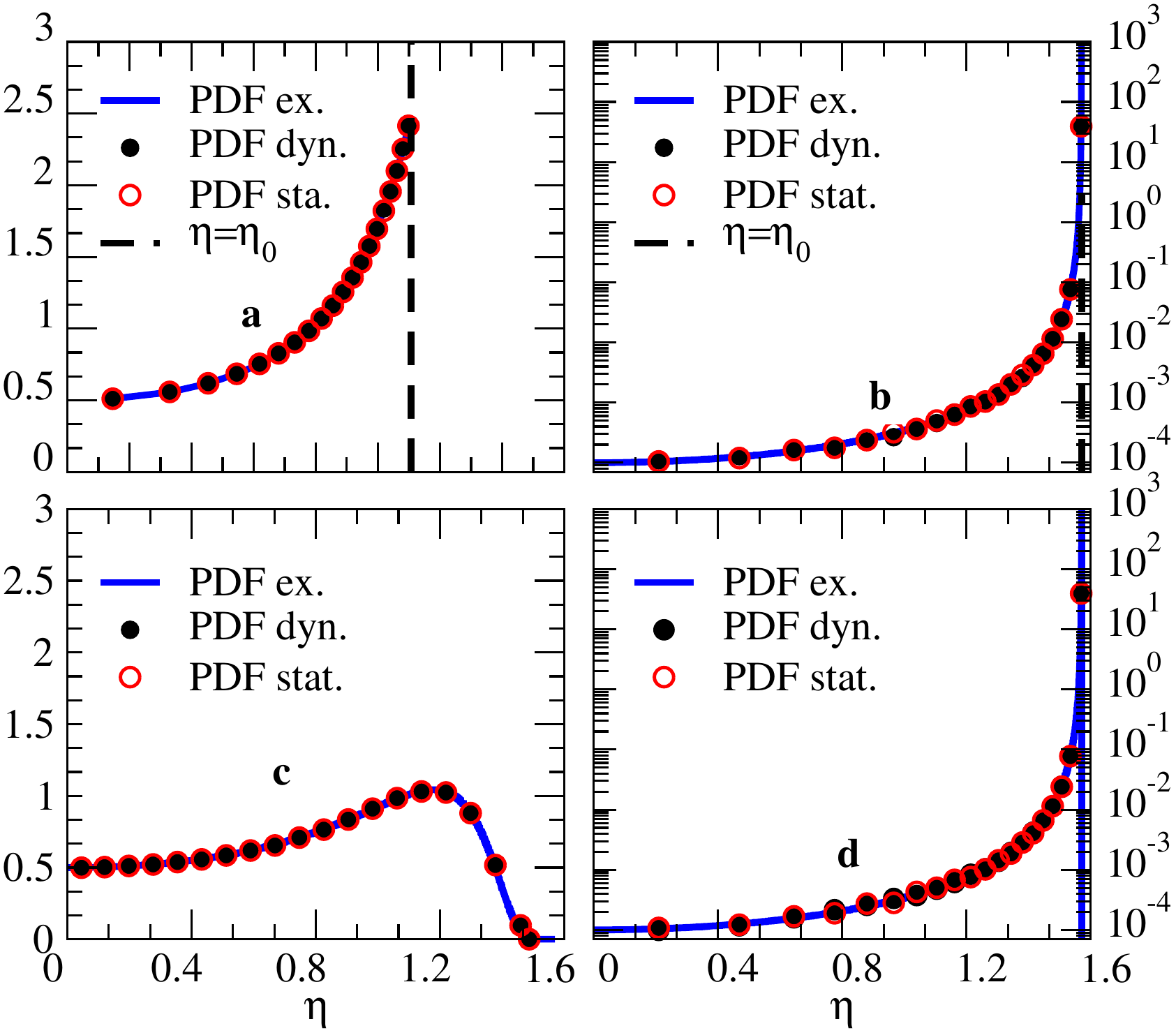}}
\caption{(Color online) Probability density functions~(PDFs) for conformal time coordinates $\eta$
in equilibrium and growing ensembles of de Sitter causal sets.
Panels~(a) and~(b) deal with the fixed-$V$ ensembles $\G_{S,V}$ and $\G_{D,V}$,
where the exact $\eta$-PDF is given by Eq.~(\ref{eq:rho(eta)-V}), the solid blue curves.
The red circles and black dots are simulation results for $\G_{S,V}$ and $\G_{D,V}$.
In~(a), the total volume is $V=2 \times 2\pi$, so that $\eta_0=\arctan2=1.107148$.
In the growing ensemble, this volume is grown in two increments $V_1=V_2=2\pi$.
In~(b), the total volume is $V=2\pi\times 10^4$, so that $\eta_0=\arctan10^4=1.570696$.
In the growing ensemble, this volume is grown in $10^4$ increments $V_1=\ldots=V_{10^4}=2\pi$.
Panels~(c) and~(d) show the corresponding results for fixed-$N$ ensembles $\G_{S,N}$ and $\G_{D,N}$,
where the exact PDF is given by Eq.~(\ref{eq:rho(eta)-N}).
In~(c), the total number of nodes is $N=2$.
In the growing ensemble, this number is grown in two increments of one node each, $N_1=N_2=1$.
In~(d), the total number of nodes is $N=10^4$.
In the growing ensemble, this number is grown in $10^4$ increments of one node each, $N_1=\ldots=N_{10^4}=1$.
The numbers of samples (PPP runs) in simulations are $S=10^6$ in~(a) and~(c), and $S=10^4$ in~(b) and~(d).
The sprinkling density (PPP rate) is $\delta=1$ everywhere.
The data for the equilibrium ensembles (red circles) almost fully overlap with the data for the growing ensembles (black dots),
both matching perfectly the theoretical predictions (blue curves).
\label{fig:1}
}
\end{figure}

The fixed-$V$ ensembles $\G_{S,V}=\G_{D,V}$ are straightforward: the $v$-PDF in them is uniform on $[0,V]$, while the $\eta$-PDF is proportional to $\sec^2\eta$:
\begin{align}
\rho(v)&=\frac{1}{V},&\text{where }&v\in[0,V],\\
\rho(\eta)&=\frac{\sec^2\eta}{\tan\eta_0},&\text{where }&\eta\in[0,\eta_0],\text{ and}\label{eq:rho(eta)-V}\\
V&=2\pi\tan\eta_0.
\end{align}
Figures~\ref{fig:1}(a,b) confirm that the $\eta$-PDF is indeed the same in $\G_{S,V}$ and $\G_{D,V}$, and equal to~(\ref{eq:rho(eta)-V}).

In the fixed-$N$ ensembles $\G_{S,N}=\G_{D,N}$, even at $N=1$, the single node can have an arbitrarily large coordinate, meaning that the corresponding distributions are defined on the whole infinite space $v\in[0,\infty)$ and $\eta\in[0,\pi/2)$. Since the distribution of the $v$-coordinate of the $i$'th node in the PPP is given by $\rho_i(v)=\Gamma_{i,\delta}(v)$, the PDF of $v$-coordinates for $N$ nodes is simply
\begin{align}
\rho(v)&=\frac{1}{N}\sum_{i=1}^N\Gamma_{i,\delta}(v)=\frac{\delta}{N}\,Q(N,\delta v),\text{ where}\\
Q(N,\delta v)&=\frac{\Gamma(N,\delta v)}{\Gamma(N)}
\end{align}
is the regularized Gamma function. The $\eta$-PDF is then
\begin{equation}\label{eq:rho(eta)-N}
\rho(\eta)=\frac{\delta}{N}\left(\sec^2\eta\right)\,Q(N,\delta\tan\eta).
\end{equation}
Figures~\ref{fig:1}(c,d) confirm that the $\eta$-PDF is indeed the same in $\G_{S,N}$ and $\G_{D,N}$, and equal to~(\ref{eq:rho(eta)-N}).

The equivalence between panels~(b) and~(d) in Fig.~\ref{fig:1} also illustrates the fact that since in the thermodynamic limit $N\to\infty$ the regularized Gamma function approaches~$1$, $Q(N,\delta v)\to1$, then
\begin{align}
\rho(v)&\to\frac{\delta}{N},\quad\text{and}\\
\rho(\eta)&\to\frac{\delta}{N}\sec^2\eta,
\end{align}
so that the fixed-$N$ ensemble $\G_{S,N}=\G_{D,N}$ with $\rho(v)=\delta/N$ becomes asymptotically identical to the fixed-$V$ ensemble $\G_{S,V}=\G_{D,V}$ with $\rho(v)=1/V$, both implementing the same PPP on the infinite half of de Sitter spacetime with $\eta>0$.

\section{Conclusion}

Almost all real networks are growing, justifying certain concerns, sometimes skepticism, about the utility of equilibrium methods in analyzing real networks. Yet the structure and dynamics of these networks turn out to be well described by network models characterized by the unusual exact equivalence between equilibrium and nonequilibrium formulations that we have proved here. These results thus provide a different perspective and further theoretical grounds for the use of powerful equilibrium methods in the analysis of real networks.

\begin{figure}
{\includegraphics[width=3.5in]{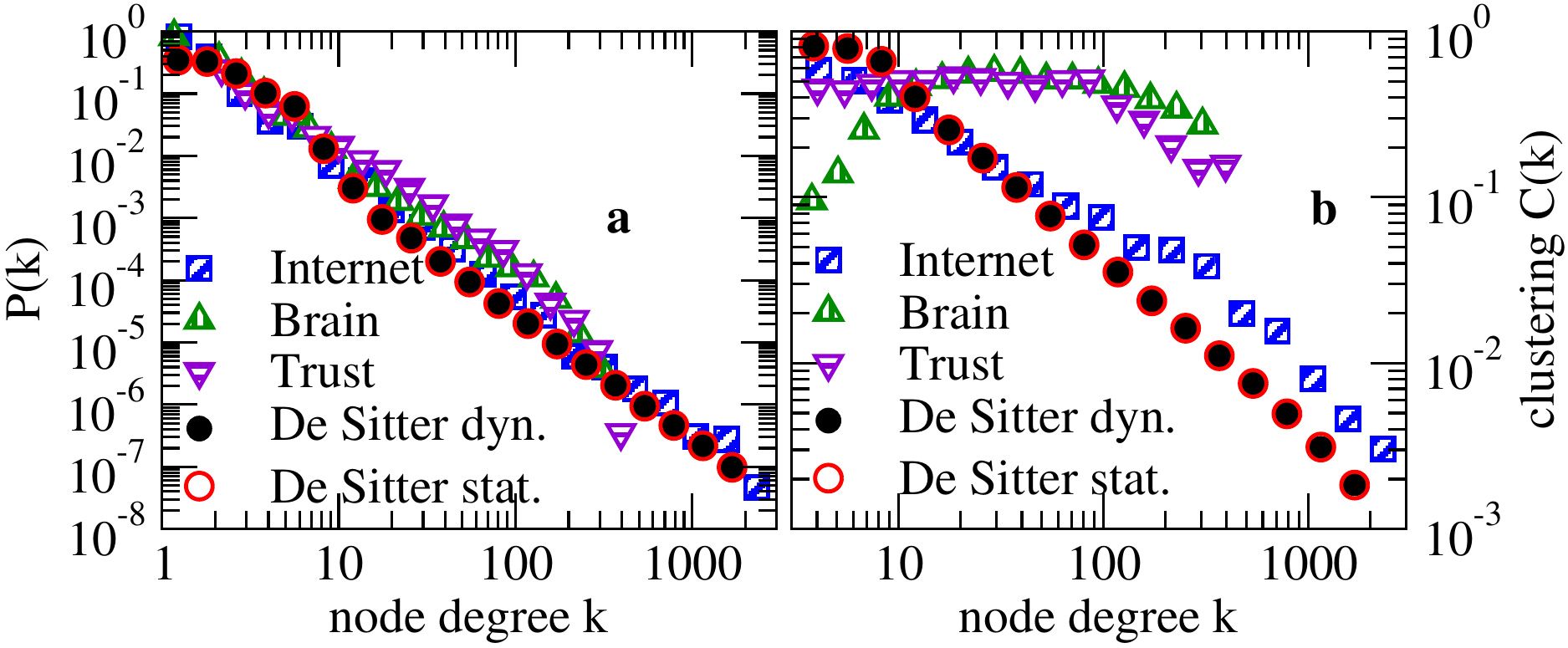}}
\caption{(Color online)
Degree distribution~(a) and clustering~(b)
in de Sitter causets $\mathcal{G}_{S,N}$ and $\mathcal{G}_{D,N}$,
$1000$ samples each of size $N=23752$,
sprinkling density $\delta=0.834751/2\pi$,
average degree $\bar{k}=4.9188$
and average clustering
$\bar{c}=0.7862$ and $\bar{c}=0.7854$, respectively, and in
some real networks: AS Internet
($N=23752$, $\bar{k}=4.9188$, $\bar{c}=0.6055$),
functional brain network
($N=23713$, $\bar{k}=6.1436$, $\bar{c}=0.1552$),
and PGP Web of Trust
($N=23797$, $\bar{k}=7.8587$, $\bar{c}=0.4816$).
\label{fig:2}}
\end{figure}

Concerning how well these strongly dual models describe real networks, we have to stress that strong duality strictly holds in de Sitter causal sets, whereas it holds only approximately in growing hyperbolic graphs in~\cite{PaBoKr11} that describe well the structure and growth dynamics of some real networks. The growth dynamics of these hyperbolic graphs becomes identical to the growth dynamics of de Sitter causets only in the thermodynamic limit~\cite{KrKi12}, and only for a specific (default) set of parameters in~\cite{PaBoKr11}. With these specific values of parameters, the model generates graphs with specific properties. These properties are: 1)~power-law degree distributions with exponent $\gamma=2$, and 2)~strongest possible clustering, i.e.\ zero temperature in~\cite{PaBoKr11}. The former property applies to many real networks, Fig.~\ref{fig:2}(a), but the latter does not, explaining differences between clustering in de Sitter causets and real networks, Fig.~\ref{fig:2}(b). However, even though clustering in real networks is not the strongest possible, it is still strong and thermodynamically stable, often even increasing slightly as the network grows~\cite{PaBoKr11}. This means that the temperature of real networks, quantifying how far their connection probability is from the step-function (zero-temperature) connection probability in causal sets, is in fact low~\cite{PaBoKr11}. In that respect the differences between de Sitter causets and real networks may be not that substantial.

The fact that strong stasis-dynamics duality holds for causal sets may be interesting from the philosophical perspective, where the problem of time finds new thematic developments in the causal set context~\cite{Wuthrich2013}. Intuitively, the finding that causal sets are strongly dual is expected because on the one hand, causal sets attempt to describe a quantization of static spacetime, while on the other hand, time is dynamic.

In general however, the exact equivalence between equilibrium and nonequilibrium systems is somewhat unusual in physics, and may even appear as a paradox. In the realm of network models for example, the required duality conditions are rarely satisfied, and do not hold in most familiar network models, such as classical random graphs, configuration model, or preferential attachment. Perhaps the most fundamental counter-example to this usual non-equivalence is the connection between equilibrium statistical mechanics (ESM) and Euclidean quantum field theory~(QFT)~\cite{McCoy2010-book}. In both cases the expression for the expected value of some observable ${\cal O}$ can be formally written as
\begin{align}
\bar{\cal O}&=\frac{1}{Z}\int_x{\cal O}\,e^{-\frac{S(x)}{h}}dx,\quad\text{where}\label{eq:O}\\
Z&=\int_xe^{-\frac{S(x)}{h}}dx
\end{align}
is the partition function. Notation triplet $S,x,h$ stands for energy, state, and $kT$ in ESM, while the same triplet in QFT is action, quantum fields, and the Planck constant. Observable ${\cal O}$ can be magnetization in ESM or scattering amplitude in QFT. This example is clearly different from the network duality that we have considered here. This network duality is the equivalence between two apparently different systems, which as we can easily prove, are actually the same because they are two different reflections of the same underlying process, a PPP. Equation~(\ref{eq:O}) describes drastically different physical systems---classical equilibrium systems and quantum nonequilibrium processes. They are not the same in any sense, and no analogy of ``an underlying PPP'' explaining~(\ref{eq:O}) is currently known, although there are some speculations~\cite{Cardy2003}. Therefore, the similarity between the ESM-QFT connection and network duality discussed here is perhaps limited to the frequent observation that the same mathematics describes two different physical phenomena.

\begin{acknowledgments}
We thank D.~Meyer, D.~Rideout, S.~Dorogovtsev, P.~Krapivsky, Z.~Toroczkai, and M.~Bogu{\~n}{\'a} for useful discussions and suggestions. This work was supported by DARPA grant No.\ HR0011-12-1-0012; NSF grants No.\ CNS-0964236 and CNS-1039646; and by Cisco Systems.
\end{acknowledgments}


\begin{thebibliography}{40}
\expandafter\ifx\csname natexlab\endcsname\relax\def\natexlab#1{#1}\fi
\expandafter\ifx\csname bibnamefont\endcsname\relax
  \def\bibnamefont#1{#1}\fi
\expandafter\ifx\csname bibfnamefont\endcsname\relax
  \def\bibfnamefont#1{#1}\fi
\expandafter\ifx\csname citenamefont\endcsname\relax
  \def\citenamefont#1{#1}\fi
\expandafter\ifx\csname url\endcsname\relax
  \def\url#1{\texttt{#1}}\fi
\expandafter\ifx\csname urlprefix\endcsname\relax\def\urlprefix{URL }\fi
\providecommand{\bibinfo}[2]{#2}
\providecommand{\eprint}[2][]{\url{#2}}

\bibitem[{\citenamefont{Dorogovtsev and Mendes}(2002)}]{DorMen02a}
\bibinfo{author}{\bibfnamefont{S.~N.} \bibnamefont{Dorogovtsev}}
  \bibnamefont{and} \bibinfo{author}{\bibfnamefont{J.~F.~F.}
  \bibnamefont{Mendes}}, \bibinfo{journal}{Adv Phys}
  \textbf{\bibinfo{volume}{51}}, \bibinfo{pages}{1079} (\bibinfo{year}{2002}).

\bibitem[{\citenamefont{Newman}(2010)}]{Newman10-book}
\bibinfo{author}{\bibfnamefont{M.~E.~J.} \bibnamefont{Newman}},
  \emph{\bibinfo{title}{{Networks: An Introduction}}}
  (\bibinfo{publisher}{Oxford University Press}, \bibinfo{address}{Oxford},
  \bibinfo{year}{2010}).

\bibitem[{\citenamefont{Solomonoff and Rapoport}(1951)}]{SoRa51}
\bibinfo{author}{\bibfnamefont{R.}~\bibnamefont{Solomonoff}} \bibnamefont{and}
  \bibinfo{author}{\bibfnamefont{A.}~\bibnamefont{Rapoport}},
  \bibinfo{journal}{B Math Biophys} \textbf{\bibinfo{volume}{13}},
  \bibinfo{pages}{107} (\bibinfo{year}{1951}).

\bibitem[{\citenamefont{Bogu\~{n}\'{a} and Pastor-Satorras}(2003)}]{BoPa03}
\bibinfo{author}{\bibfnamefont{M.}~\bibnamefont{Bogu\~{n}\'{a}}}
  \bibnamefont{and}
  \bibinfo{author}{\bibfnamefont{R.}~\bibnamefont{Pastor-Satorras}},
  \bibinfo{journal}{Phys Rev E} \textbf{\bibinfo{volume}{68}},
  \bibinfo{pages}{36112} (\bibinfo{year}{2003}).

\bibitem[{\citenamefont{Krapivsky et~al.}(2000)\citenamefont{Krapivsky, Redner,
  and Leyvraz}}]{KraReLe00}
\bibinfo{author}{\bibfnamefont{P.~L.} \bibnamefont{Krapivsky}},
  \bibinfo{author}{\bibfnamefont{S.}~\bibnamefont{Redner}}, \bibnamefont{and}
  \bibinfo{author}{\bibfnamefont{F.}~\bibnamefont{Leyvraz}},
  \bibinfo{journal}{Phys Rev Lett} \textbf{\bibinfo{volume}{85}},
  \bibinfo{pages}{4629} (\bibinfo{year}{2000}).

\bibitem[{\citenamefont{Dorogovtsev et~al.}(2000)\citenamefont{Dorogovtsev,
  Mendes, and Samukhin}}]{DoMeSa00}
\bibinfo{author}{\bibfnamefont{S.~N.} \bibnamefont{Dorogovtsev}},
  \bibinfo{author}{\bibfnamefont{J.~F.~F.} \bibnamefont{Mendes}},
  \bibnamefont{and} \bibinfo{author}{\bibfnamefont{A.~N.}
  \bibnamefont{Samukhin}}, \bibinfo{journal}{Phys Rev Lett}
  \textbf{\bibinfo{volume}{85}}, \bibinfo{pages}{4633} (\bibinfo{year}{2000}).

\bibitem[{\citenamefont{Bialas et~al.}(2003)\citenamefont{Bialas, Burda,
  Jurkiewicz, and Krzywicki}}]{BiBu03}
\bibinfo{author}{\bibfnamefont{P.}~\bibnamefont{Bialas}},
  \bibinfo{author}{\bibfnamefont{Z.}~\bibnamefont{Burda}},
  \bibinfo{author}{\bibfnamefont{J.}~\bibnamefont{Jurkiewicz}},
  \bibnamefont{and}
  \bibinfo{author}{\bibfnamefont{A.}~\bibnamefont{Krzywicki}},
  \bibinfo{journal}{Phys Rev E} \textbf{\bibinfo{volume}{67}},
  \bibinfo{pages}{66106} (\bibinfo{year}{2003}).

\bibitem[{\citenamefont{Bialas et~al.}(2005)\citenamefont{Bialas, Burda, and
  Waclaw}}]{BiBu05}
\bibinfo{author}{\bibfnamefont{P.}~\bibnamefont{Bialas}},
  \bibinfo{author}{\bibfnamefont{Z.}~\bibnamefont{Burda}}, \bibnamefont{and}
  \bibinfo{author}{\bibfnamefont{B.}~\bibnamefont{Waclaw}},
  \bibinfo{journal}{AIP Conf Proc} \textbf{\bibinfo{volume}{776}},
  \bibinfo{pages}{14} (\bibinfo{year}{2005}).

\bibitem[{\citenamefont{Park and Newman}(2004)}]{PaNe04}
\bibinfo{author}{\bibfnamefont{J.}~\bibnamefont{Park}} \bibnamefont{and}
  \bibinfo{author}{\bibfnamefont{M.~E.~J.} \bibnamefont{Newman}},
  \bibinfo{journal}{Phys Rev E} \textbf{\bibinfo{volume}{70}},
  \bibinfo{pages}{66117} (\bibinfo{year}{2004}).

\bibitem[{\citenamefont{Garlaschelli and Loffredo}(2009)}]{GaLo09}
\bibinfo{author}{\bibfnamefont{D.}~\bibnamefont{Garlaschelli}}
  \bibnamefont{and} \bibinfo{author}{\bibfnamefont{M.}~\bibnamefont{Loffredo}},
  \bibinfo{journal}{Phys Rev Lett} \textbf{\bibinfo{volume}{102}},
  \bibinfo{pages}{38701} (\bibinfo{year}{2009}).

\bibitem[{\citenamefont{Squartini and Garlaschelli}(2011)}]{SqGa11}
\bibinfo{author}{\bibfnamefont{T.}~\bibnamefont{Squartini}} \bibnamefont{and}
  \bibinfo{author}{\bibfnamefont{D.}~\bibnamefont{Garlaschelli}},
  \bibinfo{journal}{New J Phys} \textbf{\bibinfo{volume}{13}},
  \bibinfo{pages}{083001} (\bibinfo{year}{2011}).

\bibitem[{\citenamefont{Bianconi}(2008)}]{Bianconi08}
\bibinfo{author}{\bibfnamefont{G.}~\bibnamefont{Bianconi}},
  \bibinfo{journal}{Europhys Lett} \textbf{\bibinfo{volume}{81}},
  \bibinfo{pages}{28005} (\bibinfo{year}{2008}).

\bibitem[{\citenamefont{Bianconi et~al.}(2009)\citenamefont{Bianconi, Pin, and
  Marsili}}]{BiPi09}
\bibinfo{author}{\bibfnamefont{G.}~\bibnamefont{Bianconi}},
  \bibinfo{author}{\bibfnamefont{P.}~\bibnamefont{Pin}}, \bibnamefont{and}
  \bibinfo{author}{\bibfnamefont{M.}~\bibnamefont{Marsili}},
  \bibinfo{journal}{Proc Natl Acad Sci USA} \textbf{\bibinfo{volume}{106}},
  \bibinfo{pages}{11433} (\bibinfo{year}{2009}).

\bibitem[{\citenamefont{Anand and Bianconi}(2009)}]{AnBi09}
\bibinfo{author}{\bibfnamefont{K.}~\bibnamefont{Anand}} \bibnamefont{and}
  \bibinfo{author}{\bibfnamefont{G.}~\bibnamefont{Bianconi}},
  \bibinfo{journal}{Phys Rev E} \textbf{\bibinfo{volume}{80}},
  \bibinfo{pages}{045102(R)} (\bibinfo{year}{2009}).

\bibitem[{\citenamefont{Anand et~al.}(2011)\citenamefont{Anand, Bianconi, and
  Severini}}]{AnBi11}
\bibinfo{author}{\bibfnamefont{K.}~\bibnamefont{Anand}},
  \bibinfo{author}{\bibfnamefont{G.}~\bibnamefont{Bianconi}}, \bibnamefont{and}
  \bibinfo{author}{\bibfnamefont{S.}~\bibnamefont{Severini}},
  \bibinfo{journal}{Phys Rev E} \textbf{\bibinfo{volume}{83}},
  \bibinfo{pages}{036109} (\bibinfo{year}{2011}).

\bibitem[{\citenamefont{Zhao et~al.}(2011{\natexlab{a}})\citenamefont{Zhao,
  Halu, Severini, and Bianconi}}]{ZhHa11}
\bibinfo{author}{\bibfnamefont{K.}~\bibnamefont{Zhao}},
  \bibinfo{author}{\bibfnamefont{A.}~\bibnamefont{Halu}},
  \bibinfo{author}{\bibfnamefont{S.}~\bibnamefont{Severini}}, \bibnamefont{and}
  \bibinfo{author}{\bibfnamefont{G.}~\bibnamefont{Bianconi}},
  \bibinfo{journal}{Phys Rev E} \textbf{\bibinfo{volume}{84}},
  \bibinfo{pages}{066113} (\bibinfo{year}{2011}{\natexlab{a}}).

\bibitem[{\citenamefont{Zhao et~al.}(2011{\natexlab{b}})\citenamefont{Zhao,
  Karsai, and Bianconi}}]{ZhKa11}
\bibinfo{author}{\bibfnamefont{K.}~\bibnamefont{Zhao}},
  \bibinfo{author}{\bibfnamefont{M.}~\bibnamefont{Karsai}}, \bibnamefont{and}
  \bibinfo{author}{\bibfnamefont{G.}~\bibnamefont{Bianconi}},
  \bibinfo{journal}{PloS One} \textbf{\bibinfo{volume}{6}},
  \bibinfo{pages}{e28116} (\bibinfo{year}{2011}{\natexlab{b}}).

\bibitem[{\citenamefont{West et~al.}(2012)\citenamefont{West, Bianconi,
  Severini, and Teschendorff}}]{WeBi12}
\bibinfo{author}{\bibfnamefont{J.}~\bibnamefont{West}},
  \bibinfo{author}{\bibfnamefont{G.}~\bibnamefont{Bianconi}},
  \bibinfo{author}{\bibfnamefont{S.}~\bibnamefont{Severini}}, \bibnamefont{and}
  \bibinfo{author}{\bibfnamefont{A.~E.} \bibnamefont{Teschendorff}},
  \bibinfo{journal}{Sci Rep} \textbf{\bibinfo{volume}{2}}, \bibinfo{pages}{802}
  (\bibinfo{year}{2012}).

\bibitem[{\citenamefont{Penrose}(2003)}]{Penrose03-book}
\bibinfo{author}{\bibfnamefont{M.}~\bibnamefont{Penrose}},
  \emph{\bibinfo{title}{{Random Geometric Graphs}}} (\bibinfo{publisher}{Oxford
  University Press}, \bibinfo{address}{Oxford}, \bibinfo{year}{2003}).

\bibitem[{\citenamefont{Bombelli et~al.}(1987)\citenamefont{Bombelli, Lee,
  Meyer, and Sorkin}}]{BoLe87}
\bibinfo{author}{\bibfnamefont{L.}~\bibnamefont{Bombelli}},
  \bibinfo{author}{\bibfnamefont{J.}~\bibnamefont{Lee}},
  \bibinfo{author}{\bibfnamefont{D.}~\bibnamefont{Meyer}}, \bibnamefont{and}
  \bibinfo{author}{\bibfnamefont{R.}~\bibnamefont{Sorkin}},
  \bibinfo{journal}{Phys Rev Lett} \textbf{\bibinfo{volume}{59}},
  \bibinfo{pages}{521} (\bibinfo{year}{1987}).

\bibitem[{\citenamefont{Perlmutter et~al.}(1998)\citenamefont{Perlmutter,
  Aldering, Valle, Deustua, Ellis, Fabbro, Fruchter, Goldhaber, Groom, Hook
  et~al.}}]{Perlmutter98}
\bibinfo{author}{\bibfnamefont{S.}~\bibnamefont{Perlmutter}},
  \bibinfo{author}{\bibfnamefont{G.}~\bibnamefont{Aldering}},
  \bibinfo{author}{\bibfnamefont{M.~D.} \bibnamefont{Valle}},
  \bibinfo{author}{\bibfnamefont{S.}~\bibnamefont{Deustua}},
  \bibinfo{author}{\bibfnamefont{R.~S.} \bibnamefont{Ellis}},
  \bibinfo{author}{\bibfnamefont{S.}~\bibnamefont{Fabbro}},
  \bibinfo{author}{\bibfnamefont{A.}~\bibnamefont{Fruchter}},
  \bibinfo{author}{\bibfnamefont{G.}~\bibnamefont{Goldhaber}},
  \bibinfo{author}{\bibfnamefont{D.~E.} \bibnamefont{Groom}},
  \bibinfo{author}{\bibfnamefont{I.~M.} \bibnamefont{Hook}},
  \bibnamefont{et~al.}, \bibinfo{journal}{Nature}
  \textbf{\bibinfo{volume}{391}}, \bibinfo{pages}{51} (\bibinfo{year}{1998}).

\bibitem[{\citenamefont{Riess et~al.}(1998)\citenamefont{Riess, Filippenko,
  Challis, Clocchiatti, Diercks, Garnavich, Gilliland, Hogan, Jha, Kirshner
  et~al.}}]{Reiss98}
\bibinfo{author}{\bibfnamefont{A.~G.} \bibnamefont{Riess}},
  \bibinfo{author}{\bibfnamefont{A.~V.} \bibnamefont{Filippenko}},
  \bibinfo{author}{\bibfnamefont{P.}~\bibnamefont{Challis}},
  \bibinfo{author}{\bibfnamefont{A.}~\bibnamefont{Clocchiatti}},
  \bibinfo{author}{\bibfnamefont{A.}~\bibnamefont{Diercks}},
  \bibinfo{author}{\bibfnamefont{P.~M.} \bibnamefont{Garnavich}},
  \bibinfo{author}{\bibfnamefont{R.~L.} \bibnamefont{Gilliland}},
  \bibinfo{author}{\bibfnamefont{C.~J.} \bibnamefont{Hogan}},
  \bibinfo{author}{\bibfnamefont{S.}~\bibnamefont{Jha}},
  \bibinfo{author}{\bibfnamefont{R.~P.} \bibnamefont{Kirshner}},
  \bibnamefont{et~al.}, \bibinfo{journal}{Astron J}
  \textbf{\bibinfo{volume}{116}}, \bibinfo{pages}{1009} (\bibinfo{year}{1998}).

\bibitem[{\citenamefont{Krioukov et~al.}(2012)\citenamefont{Krioukov, Kitsak,
  Sinkovits, Rideout, Meyer, and Bogu\~{n}\'{a}}}]{KrKi12}
\bibinfo{author}{\bibfnamefont{D.}~\bibnamefont{Krioukov}},
  \bibinfo{author}{\bibfnamefont{M.}~\bibnamefont{Kitsak}},
  \bibinfo{author}{\bibfnamefont{R.~S.} \bibnamefont{Sinkovits}},
  \bibinfo{author}{\bibfnamefont{D.}~\bibnamefont{Rideout}},
  \bibinfo{author}{\bibfnamefont{D.}~\bibnamefont{Meyer}}, \bibnamefont{and}
  \bibinfo{author}{\bibfnamefont{M.}~\bibnamefont{Bogu\~{n}\'{a}}},
  \bibinfo{journal}{Sci Rep} \textbf{\bibinfo{volume}{2}}, \bibinfo{pages}{793}
  (\bibinfo{year}{2012}).

\bibitem[{\citenamefont{Papadopoulos et~al.}(2012)\citenamefont{Papadopoulos,
  Kitsak, Serrano, Bogu\~{n}\'{a}, and Krioukov}}]{PaBoKr11}
\bibinfo{author}{\bibfnamefont{F.}~\bibnamefont{Papadopoulos}},
  \bibinfo{author}{\bibfnamefont{M.}~\bibnamefont{Kitsak}},
  \bibinfo{author}{\bibfnamefont{M.~A.} \bibnamefont{Serrano}},
  \bibinfo{author}{\bibfnamefont{M.}~\bibnamefont{Bogu\~{n}\'{a}}},
  \bibnamefont{and} \bibinfo{author}{\bibfnamefont{D.}~\bibnamefont{Krioukov}},
  \bibinfo{journal}{Nature} \textbf{\bibinfo{volume}{489}},
  \bibinfo{pages}{537} (\bibinfo{year}{2012}).

\bibitem[{\citenamefont{Harremoes}(2001)}]{Harremoes2001}
\bibinfo{author}{\bibfnamefont{P.}~\bibnamefont{Harremoes}},
  \bibinfo{journal}{IEEE T Inform Theory} \textbf{\bibinfo{volume}{47}},
  \bibinfo{pages}{2039} (\bibinfo{year}{2001}).

\bibitem[{\citenamefont{Chung and Lu}(2002)}]{ChLu02}
\bibinfo{author}{\bibfnamefont{F.}~\bibnamefont{Chung}} \bibnamefont{and}
  \bibinfo{author}{\bibfnamefont{L.}~\bibnamefont{Lu}}, \bibinfo{journal}{Ann
  Comb} \textbf{\bibinfo{volume}{6}}, \bibinfo{pages}{125}
  (\bibinfo{year}{2002}).

\bibitem[{\citenamefont{Colomer-de Simon and Bogu\~{n}\'{a}}(2012)}]{CoBo12}
\bibinfo{author}{\bibfnamefont{P.}~\bibnamefont{Colomer-de Simon}}
  \bibnamefont{and}
  \bibinfo{author}{\bibfnamefont{M.}~\bibnamefont{Bogu\~{n}\'{a}}},
  \bibinfo{journal}{Phys Rev E} \textbf{\bibinfo{volume}{86}},
  \bibinfo{pages}{026120} (\bibinfo{year}{2012}).

\bibitem[{\citenamefont{Coolen et~al.}(2009)\citenamefont{Coolen, Martino, and
  Annibale}}]{CoMa09}
\bibinfo{author}{\bibfnamefont{A.~C.~C.} \bibnamefont{Coolen}},
  \bibinfo{author}{\bibfnamefont{A.}~\bibnamefont{Martino}}, \bibnamefont{and}
  \bibinfo{author}{\bibfnamefont{A.}~\bibnamefont{Annibale}},
  \bibinfo{journal}{J Stat Phys} \textbf{\bibinfo{volume}{136}},
  \bibinfo{pages}{1035} (\bibinfo{year}{2009}).

\bibitem[{\citenamefont{Annibale et~al.}(2009)\citenamefont{Annibale, Coolen,
  Fernandes, Fraternali, and Kleinjung}}]{AnCo09}
\bibinfo{author}{\bibfnamefont{A.}~\bibnamefont{Annibale}},
  \bibinfo{author}{\bibfnamefont{A.~C.~C.} \bibnamefont{Coolen}},
  \bibinfo{author}{\bibfnamefont{L.}~\bibnamefont{Fernandes}},
  \bibinfo{author}{\bibfnamefont{F.}~\bibnamefont{Fraternali}},
  \bibnamefont{and}
  \bibinfo{author}{\bibfnamefont{J.}~\bibnamefont{Kleinjung}},
  \bibinfo{journal}{J Phys A-Math Gen} \textbf{\bibinfo{volume}{42}},
  \bibinfo{pages}{485001} (\bibinfo{year}{2009}).

\bibitem[{\citenamefont{{Del Genio} et~al.}(2010)\citenamefont{{Del Genio},
  Kim, Toroczkai, and Bassler}}]{DeKi10}
\bibinfo{author}{\bibfnamefont{C.~I.} \bibnamefont{{Del Genio}}},
  \bibinfo{author}{\bibfnamefont{H.}~\bibnamefont{Kim}},
  \bibinfo{author}{\bibfnamefont{Z.}~\bibnamefont{Toroczkai}},
  \bibnamefont{and} \bibinfo{author}{\bibfnamefont{K.~E.}
  \bibnamefont{Bassler}}, \bibinfo{journal}{PloS one}
  \textbf{\bibinfo{volume}{5}}, \bibinfo{pages}{e10012} (\bibinfo{year}{2010}).

\bibitem[{\citenamefont{Krioukov et~al.}(2010)\citenamefont{Krioukov,
  Papadopoulos, Kitsak, Vahdat, and Bogu\~{n}\'{a}}}]{KrPa10}
\bibinfo{author}{\bibfnamefont{D.}~\bibnamefont{Krioukov}},
  \bibinfo{author}{\bibfnamefont{F.}~\bibnamefont{Papadopoulos}},
  \bibinfo{author}{\bibfnamefont{M.}~\bibnamefont{Kitsak}},
  \bibinfo{author}{\bibfnamefont{A.}~\bibnamefont{Vahdat}}, \bibnamefont{and}
  \bibinfo{author}{\bibfnamefont{M.}~\bibnamefont{Bogu\~{n}\'{a}}},
  \bibinfo{journal}{Phys Rev E} \textbf{\bibinfo{volume}{82}},
  \bibinfo{pages}{36106} (\bibinfo{year}{2010}).

\bibitem[{\citenamefont{Hawking et~al.}(1976)\citenamefont{Hawking, King, and
  McCarthy}}]{HaKi76}
\bibinfo{author}{\bibfnamefont{S.~W.} \bibnamefont{Hawking}},
  \bibinfo{author}{\bibfnamefont{A.~R.} \bibnamefont{King}}, \bibnamefont{and}
  \bibinfo{author}{\bibfnamefont{P.~J.} \bibnamefont{McCarthy}},
  \bibinfo{journal}{J Math Phys} \textbf{\bibinfo{volume}{17}},
  \bibinfo{pages}{174} (\bibinfo{year}{1976}).

\bibitem[{\citenamefont{Malament}(1977)}]{Malament77}
\bibinfo{author}{\bibfnamefont{D.~B.} \bibnamefont{Malament}},
  \bibinfo{journal}{J Math Phys} \textbf{\bibinfo{volume}{18}},
  \bibinfo{pages}{1399} (\bibinfo{year}{1977}).

\bibitem[{\citenamefont{Rideout and Sorkin}(1999)}]{RiSo99}
\bibinfo{author}{\bibfnamefont{D.}~\bibnamefont{Rideout}} \bibnamefont{and}
  \bibinfo{author}{\bibfnamefont{R.}~\bibnamefont{Sorkin}},
  \bibinfo{journal}{Phys Rev D} \textbf{\bibinfo{volume}{61}},
  \bibinfo{pages}{024002} (\bibinfo{year}{1999}).

\bibitem[{\citenamefont{Hawking and Ellis}(1975)}]{HawkingEllis1975}
\bibinfo{author}{\bibfnamefont{S.~W.} \bibnamefont{Hawking}} \bibnamefont{and}
  \bibinfo{author}{\bibfnamefont{G.~F.~R.} \bibnamefont{Ellis}},
  \emph{\bibinfo{title}{{The Large Scale Structure of Space-Time}}}
  (\bibinfo{publisher}{Cambridge University Press},
  \bibinfo{address}{Cambridge}, \bibinfo{year}{1975}).

\bibitem[{\citenamefont{Snyder and Miller}(2011)}]{Snyder2011-book}
\bibinfo{author}{\bibfnamefont{D.~L.} \bibnamefont{Snyder}} \bibnamefont{and}
  \bibinfo{author}{\bibfnamefont{M.~I.} \bibnamefont{Miller}},
  \emph{\bibinfo{title}{{Random Point Processes in Time and Space}}}
  (\bibinfo{publisher}{Springer}, \bibinfo{address}{New York},
  \bibinfo{year}{2011}).

\bibitem[{\citenamefont{{H. David} and {H.
  Nagaraja}}(2003)}]{OrderStatistics03}
\bibinfo{author}{\bibnamefont{{H. David}}} \bibnamefont{and}
  \bibinfo{author}{\bibnamefont{{H. Nagaraja}}}, \emph{\bibinfo{title}{{Order
  Statistics}}} (\bibinfo{publisher}{Wiley-Interscience}, \bibinfo{address}{New
  York}, \bibinfo{year}{2003}).

\bibitem[{\citenamefont{W\"{u}thrich}(2013)}]{Wuthrich2013}
\bibinfo{author}{\bibfnamefont{C.}~\bibnamefont{W\"{u}thrich}},
  \bibinfo{journal}{J Gen Philos Sci} \textbf{\bibinfo{volume}{43}},
  \bibinfo{pages}{223} (\bibinfo{year}{2013}).

\bibitem[{\citenamefont{McCoy}(2010)}]{McCoy2010-book}
\bibinfo{author}{\bibfnamefont{B.~M.} \bibnamefont{McCoy}},
  \emph{\bibinfo{title}{{Advanced Statistical Mechanics}}}
  (\bibinfo{publisher}{Oxford University Press}, \bibinfo{address}{Oxford},
  \bibinfo{year}{2010}).

\bibitem[{\citenamefont{Cardy}(2003)}]{Cardy2003}
\bibinfo{author}{\bibfnamefont{J.}~\bibnamefont{Cardy}}, \bibinfo{journal}{Ann
  Henri Poincar\'{e}} \textbf{\bibinfo{volume}{4}}, \bibinfo{pages}{371}
  (\bibinfo{year}{2003}).

\end{thebibliography}
\end{document}